%% file: main_en.tex
\documentclass[11pt,a4paper]{article}
\usepackage[T1]{fontenc}
\usepackage[utf8]{inputenc}
\usepackage{lmodern}
\usepackage{amsmath,amssymb}
\usepackage{booktabs}
\usepackage{graphicx}
\usepackage{pgfplots}
\pgfplotsset{compat=1.18}
\usepgfplotslibrary{fillbetween}
\usepackage[margin=25mm]{geometry}
\usepackage{tikz}
\usetikzlibrary{arrows.meta,positioning,fit,backgrounds,calc}
\usepackage[hidelinks]{hyperref}
\usepackage[backend=biber,style=nature]{biblatex}
\renewcommand*{\textcite}[1]{\citeauthor{#1}\autocite{#1}}

\input{figs/figstyle}
\newcommand{\PrC}{\Pr(C\mid T)}
\newcommand{\PrNC}{\Pr(C\mid\neg T)}
\newcommand{\LR}{\Lambda}
\title{Phantom Evidence\\[4pt]\large How and Why Generative AI Manufactures False Positives in Science}
\author{%
  Yukiyasu Kamitani$^{1,2}$\thanks{Correspondence: \texttt{kamitani@i.kyoto-u.ac.jp}}\quad
  Ken Shirakawa$^{2}$\\[8pt]
  {\normalsize $^{1}$Graduate School of Informatics, Kyoto University}\\[2pt]
  {\normalsize $^{2}$ATR Computational Neuroscience Laboratories}}
\date{}
\begin{document}
\maketitle
\begin{abstract}
\noindent
Four centuries ago Francis Bacon warned against the anticipations of nature, hasty generalization that wins assent on a few facts, and set against it the \textit{table of absence}: checking that a property fails to appear where it should not. The demand was that looking convincing should not, on its own, count as evidence. Science has professed that demand ever since, while in practice letting persuasiveness do the work of evidence. It could be let to do so because making something persuasive was itself hard. Generative AI removes that difficulty, and an old error returns on a scale and at a speed it never had before. We locate the problem not in evidence growing weaker but in how surprise is counted. An observer marvels at a convincing output as a single point hit among a vast range of possibilities, yet what a system can actually reach is a small part of that range. The gap between the breadth imagined and the narrowness actually reached is what we call phantom evidence, and we formalize it as one quantity that also absorbs the trial and error and the data leakage a research process adds. Three things follow. Higher resolution and greater fluency add no evidence. The evidence a single result can carry has a ceiling that neither polishing the output nor letting a generative system grade itself can exceed. And the fraction of published findings that are true falls back to what it was before anything was observed. The prescription lies in the same place: genuinely widen what a system can reach, and measure whether convincing outputs still appear when the target is absent. Bacon's table of absence, in the language of probability, is a single line: measure the probability of looking convincing when the target is absent. In a world where the persuasive has become cheap, the credibility of science rests not on more convincing outputs but on procedures that show they could not have arisen by chance.

\end{abstract}

\section{The Problem: Looking Convincing versus Being True}
Trust in a scientific claim ultimately rests on a single inference: ``this looks convincing, therefore it is probably true.'' That inference is valid only insofar as looking convincing is correlated with being true. Sever the correlation and looking convincing ceases to be evidence and becomes ornament.

Generative AI selectively destroys this correlation. Large language models and image generators are optimized to produce outputs that a human judge finds plausible. What they are not optimized for, at least not directly, is that the output be true of the world. The two often correlate loosely, but the correlation is a by-product rather than something the training objective guarantees. The decoupling is already measurable in controlled settings. \textcite{gao2023} showed that blinded reviewers mistook roughly a third (32\%) of ChatGPT-generated scientific abstracts for genuine ones, even as a dedicated AI detector identified them with high accuracy. The human judgment ``this looks convincing'' can thus diverge systematically from whether the abstract reports a sound study. Being detectable is itself evidence of the divergence, but it is not a resolution of it (Section~\ref{sec:objections}). The illusion of understanding described by \textcite{messeri2024}, in which fluent output produces a feeling of having grasped something beyond the actual soundness of its content, is another face of the same divergence.

We call this persuasion inflation. Just as an expanded money supply dilutes the value of a currency, the ability to mass-produce convincing artifacts cheaply dilutes the evidential value per unit of the signal ``persuasiveness.'' The problem is not any individual fake; it is the debasement of the signal itself.

To make this debasement precise, we define two propositions about a research artifact under evaluation (a model output, a reconstructed image, a generated proof, a reported effect).
\begin{itemize}
  \item $T$ (target): the artifact actually achieves its claimed target.
  \item $C$ (convincing): the artifact is judged convincing, or plausibly correct, under our evaluation procedure.
\end{itemize}
What we can observe is $C$, not $T$. Therein lies the whole problem. Note that $T$ is not confined to ``the truth of a hypothesis''; the carrier of the attribute is ours to choose, and the same structure extends isomorphically to at least three layers, namely the individual prediction (target fidelity), the method (validity), and the claim (truth and reproducibility). In what follows we take the prediction layer as our base case, returning to the differences among the three layers in Section~\ref{sec:ppv}, where the reproducibility crisis is discussed.

\begin{figure}[t]
\centering
\input{figs/fig_threecases_en.tikz}
\caption{One mechanism, three faces. Clever Hans, the overestimation of large language model ability, and spurious visual image reconstruction are superficially unrelated, yet in each the nominal space $N$ is vast while the effective number of options $k_{\mathrm{eff}}$ the system actually spans is small. This paper formalizes the gap $\Delta=\log(N/k_{\mathrm{eff}})$ as phantom evidence ($N$ and $k_{\mathrm{eff}}$ are defined in Section~\ref{sec:phantom}).}
\label{fig:threecases}
\end{figure}
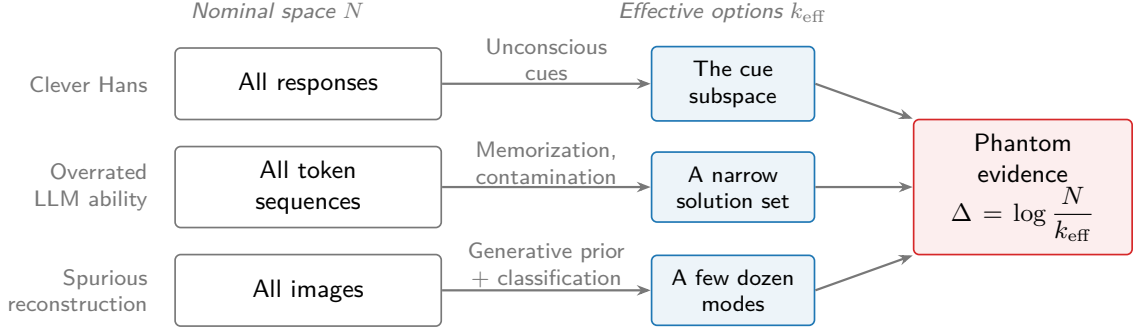

This divergence is no idiosyncrasy of one field. Consider three examples in chronological order (Figure~\ref{fig:threecases}). In the early twentieth century a horse called Clever Hans, which appeared to compute by tapping its hoof, became famous; what the horse was reading was the questioner's unconscious bodily cues~\autocite{pfungst1911}, and the same trap is found in today's machine-learning classifiers~\autocite{lapuschkin2019}. In current evaluations of large language models, claims that a model ``can reason'' or ``passes a professional examination'' are supported by correct answers, but those answers can arise equally from genuine capability and from memorization, shortcut learning~\autocite{geirhos2020shortcut} and benchmark contamination; \textcite{schaeffer2023} show that apparent emergent abilities can be an artifact of the chosen evaluation metric. In spurious reconstruction of visual images, a diffusion model's generative prior together with classification into trained categories concentrates outputs on a few typical patterns, so a photorealistic image can appear even when the brain contributes almost nothing~\autocite{shirakawa2025}.

The three are superficially unrelated but share one skeleton. The observer marvels that ``a single point was hit within a vast space,'' while the range the system actually reaches is far narrower. It is this gap between the vastness of the nominal space and the smallness of the effective range that this paper condenses into a single quantity and makes rigorous under the name phantom evidence. Generative AI mass-produces that divergence at near-zero marginal cost. The question itself is not new. Four centuries ago, in the \textit{Novum Organum}~\autocite{bacon1620}, Francis Bacon warned against the anticipations of nature (\textit{Anticipationes Naturae}), the habit of generalizing hastily from a few facts and winning assent on that basis, and set against it the \textit{table of absence} and exclusion: the operation of checking that a property fails to appear where it should not. Our claim is that this demand, put in the language of probability, takes the form \emph{measure the denominator}, and that generative AI is precisely a device for evading it (we return to this extension, under the name \textit{Idola Machinae}, in the conclusion). In what follows we show which quantity we formalize (Sections~\ref{sec:lr}--\ref{sec:phantom}), where and why generative AI produces false positives (Sections~\ref{sec:mechanisms}--\ref{sec:sites}), and what prescription follows (Section~\ref{sec:prescription}), before examining where the framework holds and where it weakens (Section~\ref{sec:scope}) and answering the objections it invites (Section~\ref{sec:objections}).

\section{Evidence Lives in the Denominator: The Likelihood Ratio at the Core}\label{sec:lr}
To diagnose the divergence we must write down, as a quantity, how far an observation moves belief. Bayes's theorem splits our post-observation belief into a prior and an evidential factor:
\begin{equation}
\underbrace{\frac{\Pr(T \mid C)}{\Pr(\neg T \mid C)}}_{\text{posterior odds}}
  = \underbrace{\frac{\Pr(T)}{\Pr(\neg T)}}_{\text{prior odds}}
  \times \underbrace{\frac{\PrC}{\PrNC}}_{\LR}.
\end{equation}
The prior odds are a subjective term derived from a field's background knowledge. The very force with which the observation $C$ moves belief, namely the likelihood ratio or Bayes factor, is
\begin{equation}
\boxed{\;\LR \;=\; \dfrac{\PrC}{\PrNC}\;}
\end{equation}
and evidential value lives in $\LR$, or, measured as information, in $\log\LR$. This $\log\LR$ is what \textcite{good1950} called the weight of evidence; the view that evidence just is a likelihood ratio is developed at length by \textcite{royall1997}. Writing the per-outcome likelihood ratio as $L(X)=\Pr(X\mid T)/\Pr(X\mid\neg T)$ for $X\in\{C,\neg C\}$, so that $L(C)=\LR$, its expectation under $T$, $\mathbb{E}_{X\mid T}[\log L(X)]$, equals the Kullback--Leibler discrimination information~\autocite{kl1951} between the conditional distributions. (In the text $\log$ is base $10$, Good's ban; figures use bits, $\log_2$, for legibility; the extreme-value and continuous appendices use natural logarithms. The conversion is 1 ban $=\log_2 10\approx3.32$ bits.) For observations that are conditionally independent both under $T$ and under $\neg T$, the ratios multiply and $\log\LR$ adds, so evidence is an additive quantity on a single scale.

Independence, however, is a strong assumption. Hits that share the same model, the same shortcut or the same contamination are mutually correlated, so the evidence carried by $n$ positive cases collapses to an effective count $n_{\mathrm{eff}}$ and saturates (Appendix~\ref{sec:subadditivity}). This is why stacking a gallery of generated artifacts does not make the evidence proportional to their number.

Note also that what $\log\LR$ measures is the diagnosticity of evidence, a quantity that scores an observation with the hypothesis held fixed; it does not track whether that observation was used in constructing or selecting the hypothesis. A large $\log\LR$ therefore does not by itself mean that the claim passed a severe test. Severity in the sense of \textcite{mayo2018} concerns a different property, the degree to which the observation would not have come out this way had the hypothesis been false, and only when $\log\LR$ is reported together with the selection ledger $m_{\mathrm{eff}}$ (Section~\ref{sec:mechanisms}) can one speak about severity as well.

From this follows the point at the core of this paper. The magnitude of $\LR$ is governed by the denominator $\PrNC$. The numerator $\PrC$, the probability that a genuine artifact looks convincing, tends to saturate near $1$ under sufficient power, so what moves $\LR$ is almost entirely the denominator. Once $\PrNC$ approaches the numerator, $\LR\to1$ and $\log\LR\to0$. Observing $C$ then carries no evidence at all, no matter how many positive cases are amassed. This is the literal sense of ``evidence lives in the denominator.'' We call this limit denominator collapse, meaning the collapse of $\LR$ that the swelling of the denominator produces, and we locate the threat of generative AI precisely there. Where we need to name the cause rather than the effect we speak of denominator inflation.

Polishing cannot add evidence. That the denominator imposes a ceiling has an information-theoretic warrant. Information about the target $T$ enters the pipeline through one gate only: the primary data, the record that is in physical contact with the target. Everything downstream is a rendering of that record, an exhibit placed before the observer (a reconstructed image, an aggregate score or a gallery of successes, a figure or a claim), on which the judgment $C$ of whether it looks convincing is made. So long as this flow forms a one-way chain, $T\to\text{primary data}\to\text{exhibit}\to C$ --- that is, so long as the exhibit is a function of the primary data alone and the judgment $C$ a (possibly stochastic) function of the exhibit alone ($\text{exhibit}\perp T\mid\text{primary data}$ and $C\perp T\mid\text{exhibit}$) --- the data-processing inequality gives the last two inequalities in
\begin{equation}
H(T)\;\ge\;I(T;\text{primary data})\;\ge\;I(T;\text{exhibit})\;\ge\;I(T;C)
\end{equation}
while the first is simply $I\le H$.\footnote{Here $I$ is an expected information, averaged over the distribution of $T$, and is a different quantity from the weight of evidence $\log\LR$ carried by a particular observation. If $T$ is binary then $I(T;C)\le H(T)\le1$ bit, whereas $\log\LR$ is unbounded. Where we speak below of a ceiling on the effective evidence we mean the bound $\log\LR_{\mathrm{actual}}\le\log k_{\mathrm{eff}}$ (Section~\ref{sec:prescription}).} In other words, any processing, polishing, or improvement of appearance carried out after the primary data have been recorded cannot add a single bit of information about the target; it is merely a function placed downstream in the chain. What downstream work can do is not to add information but to avoid losing it.

Repairing a mapping that discards information (output dimension collapse, Section~\ref{sec:sites}) brings $I(T;\text{exhibit})$ closer to its ceiling $I(T;\text{primary data})$, it does not raise the ceiling (Section~\ref{sec:prescription}, direction 1). Raising photorealism and having a generative system itself do the grading (LLM-as-judge, Section~\ref{sec:prescription}(5)) are both downstream operations, provided the judge touches no independent information beyond the exhibit (on oracles, meaning independent verifiers that a fake cannot cheaply pass, see Section~\ref{sec:prescription}(5)). Neither adds evidence.

Conversely, so long as the chain holds, the inequality holds; it can break only when information about $T$ enters from the side, bypassing the primary data. Both the leakage of Section~\ref{sec:mechanisms} and selection by someone who knows the target are such side routes: in leakage the evaluation ground truth enters the exhibit through preprocessing or the metric, and in selection the rule for choosing a path itself depends on $T$. Note that merely trying many paths and reporting the best, without knowing the ground truth, leaves the chain intact; there the inequality does not break, but $\PrNC$ rises and so $I(T;C)$ and $\log\LR$ fall. Side route or in-chain degradation, either way the evidence shrinks.

(Numerator saturation is an approximation, not a universal law; on where it fails, see Section~\ref{sec:scope}. The core claim that the denominator imposes the ceiling does not depend on saturation.)

\section{Phantom Evidence: The Vast Nominal and the Narrow Effective}\label{sec:phantom}

Denominator collapse is one limit of a more general phenomenon. The observer regards a convincing output as a single point selected from a nominal candidate set of size $N$, the set spanned by combinations of pixels, tokens and hypotheses. The rarity of that point is then read as ``$1/N$ by chance,'' and that reading underlies the perceived likelihood ratio $\LR_{\mathrm{perceived}}=\PrC/(1/N)$. Hereafter $N$ denotes the number of candidates the observer assumes: in free generation the number of cells fixed by the resolution, and in forced-choice identification the number of alternatives including the distractors (Section~\ref{sec:sites}). But the generator's output distribution $p_g$ concentrates on the convincing region, so the true denominator $\PrNC$, the probability of a convincing output even when the target is absent, is far larger than the $1/N$ the observer assumes. In this section we treat only one of the factors that inflate the denominator, namely the concentration of the generator's output, and add the selection and leakage that arise from the research process in the next section. Writing that bare denominator as $\gamma$, we define the effective number of options for the judgment $C$ operationally as
\begin{equation}
k_{\mathrm{eff}} \;:=\; \frac{1}{\gamma}
\end{equation}
that is, the reciprocal of ``the probability of $C$ even under $\neg T$ from the generator's concentration alone.'' The actual likelihood ratio is then $\LR_{\mathrm{actual}}=\PrC/\gamma=\PrC\cdot k_{\mathrm{eff}}$, and we call its gap from $\LR_{\mathrm{perceived}}$ phantom evidence:
\begin{equation}
\Delta \;=\; \log\LR_{\mathrm{perceived}}-\log\LR_{\mathrm{actual}}
\;=\; \log\big(N\gamma\big)
\;=\; \log\frac{N}{k_{\mathrm{eff}}}.
\end{equation}
The selection- and leakage-inclusive effective denominator $\PrNC\ (\ge\gamma)$, and the corresponding total $\Delta_{\mathrm{total}}=\log(N\,\PrNC)$, are derived in Section~\ref{sec:mechanisms}. Hereafter $k_{\mathrm{eff}}$ always denotes the bare value $1/\gamma$, and the reciprocal of the effective denominator as measured by a negative control is written separately as $k_{\mathrm{eff}}^{\mathrm{obs}}:=1/\PrNC\ (\le k_{\mathrm{eff}})$.

$\Delta$ is the amount by which the observer overestimates the weight of evidence (in bans, if the base is $10$). Under an idealization that averages the generator's output over the whole distribution, this quantity equals the divergence from the uniform distribution $u_N$, namely $D_{\mathrm{KL}}(p_g\|u_N)=\log N-H(p_g)$, where $k_{\mathrm{eff}}$ corresponds to the generator's entropic effective number of modes $10^{H_{10}(p_g)}$. (The two agree exactly only under an idealization that is uniform over the effective modes; in general they differ by the choice of R\'enyi order. We take the operational definition $k_{\mathrm{eff}}=1/\gamma$ as canonical throughout. It is the reciprocal of the $p_g$-mass of the accepting region, and it coincides with a collision-type count of order R\'enyi-2 when the target is itself typical of $p_g$, which is the regime that birthday-paradox estimators probe~\autocite{arora2018}. The entropic and covering-number versions used below and in Appendix~\ref{sec:continuous} are the R\'enyi-1 and R\'enyi-0 readings and are larger. We flag that this is the most favourable of the three readings to our own thesis, since the smallest $k_{\mathrm{eff}}$ yields the largest $\Delta$ and the lowest ceiling; a reader who prefers the entropic reading should scale our $\Delta$ down accordingly. On the ordering of effective counts by R\'enyi order see \textcite{hill1973}.) It measures how concentrated $p_g$ is relative to uniform, which is the effective poverty of the output.

\begin{figure}[t]
\centering
\input{figs/fig_boxes_en.tikz}
\caption{The intuition of phantom evidence: nominal $N$ against effective $k_{\mathrm{eff}}$. For the same hit (red), the number of boxes over which surprise is counted differs. Left: the nominal space the observer assumes (the faint whole space of size $N$, felt as ``$1/N$ by chance''). Right: one cell of that space magnified by a factor of $N/k_{\mathrm{eff}}=2\times10^4$, which is the effective space the generator actually reaches (the dense modes with $k_{\mathrm{eff}}\ll N$, which should really be counted as $1/k_{\mathrm{eff}}$). The effective space does not lie outside the nominal space; it is only a tiny part of its interior. The gap $\log(N/k_{\mathrm{eff}})$ between the two is phantom evidence.}
\label{fig:boxes}
\end{figure}
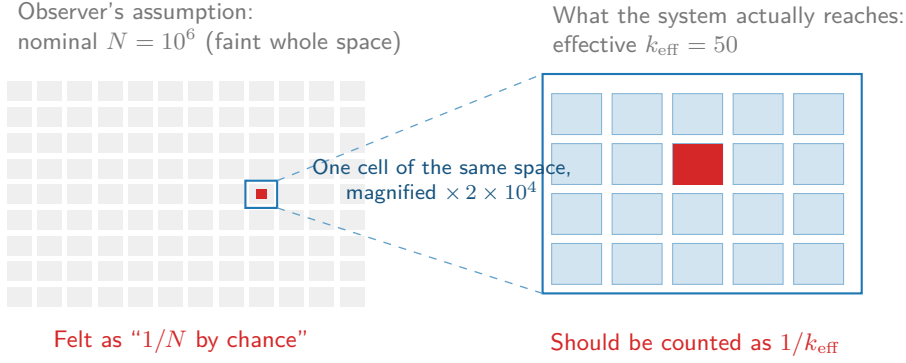

In the limit $\LR_{\mathrm{actual}}\to1$ the actual evidence vanishes and nearly the entire amount the observer feels remains as phantom evidence. High resolution, high dimensionality and fluency inflate $N$ exponentially without raising $k_{\mathrm{eff}}$, so only phantom evidence expands (Figure~\ref{fig:boxes}).

The three examples of the opening (Figure~\ref{fig:threecases}) are all manifestations of this single quantity $\Delta$. The observer believes $\log N$ to be the evidence, while the effective amount is only $\log k_{\mathrm{eff}}$. In what follows we develop this quantity through the mechanisms (Section~\ref{sec:mechanisms}), the posterior probability (Section~\ref{sec:ppv}) and a worked example (Section~\ref{sec:sites}). All figures share one set of illustrative parameter values, which we call the running example (it is a parameter set, not to be confused with the worked example of Section~\ref{sec:sites}, which is a research case): nominal $N=10^6$, effective $k_{\mathrm{eff}}=50$ (hence $N/k_{\mathrm{eff}}=2\times10^4$), selection multiplicity $m_{\mathrm{eff}}=20$, and power $q=\PrC=0.9$. The worked example of Section~\ref{sec:sites} takes separate values keyed to real data.

It should be said plainly that the bare $k_{\mathrm{eff}}$ is not a new quantity to measure. The effective support size of a generator has been estimated directly by birthday-paradox tests~\autocite{arora2018}, and the coverage of a generator's output relative to the data distribution is what the recall side of precision-and-recall metrics reports~\autocite{kynkaanniemi2019}. What is not available off the shelf is the procedure-inclusive $k_{\mathrm{eff}}^{\mathrm{obs}}=1/\PrNC$, measured by running the whole analysis, selection rule included, on a negative control. Nor is the ledger: this paper puts the bare count, the selection multiplicity and the leakage into one expression that says how much evidence a reader is over-crediting, and ties that expression to the design of the control that would measure it.

\section{Mechanisms: Why Generative Systems Inflate the Denominator}\label{sec:mechanisms}
That the denominator $\PrNC$ swells is not an accidental malfunction; it is written into the objective function of generative systems themselves. A language model is trained to approximate the distribution of text humans judge fluent and plausible, and reinforcement learning from human feedback (RLHF) skews the optimization target further toward looking desirable to an evaluator. Unless verification is built into the training loop, no constraint that ``the output be true of the world'' enters. The generator therefore systematically acquires the ability to emit convincing outputs even in regions where the target is absent. The word ``hallucination'' pathologizes and individualizes this phenomenon, but structurally it is nothing other than normal operation exactly as specified by the objective function: the expected rise of $\PrNC$.

The generator is not the only source of denominator inflation. The research process itself has pushed it up through two isomorphic routes since long before generative AI. One is selection: reporting the best among branches of preprocessing, metrics and subsets (researcher degrees of freedom, p-hacking~\autocite{simmons2011}; the garden of forking paths~\autocite{gelman2014garden}). The loss of validity that this adaptivity causes has been given a quantitative theory in the adaptive data analysis literature, where the reusable holdout shows how a held-out set can be queried many times over without its guarantees expiring~\autocite{dwork2015}. The other is leakage: the double dipping and circular analysis in which evaluation data contaminates feature selection, normalization or training~\autocite{kriegeskorte2009}. Consider fishing. Fish once, and a lucky big catch is rare (denominator $\approx1/k_{\mathrm{eff}}$). But fish dozens of times, report only the largest, and a lucky big catch becomes routine (selection). Seed the pond with the answers beforehand, and the catch is no longer evidence of skill (leakage). Generative AI automates and accelerates both routes at near-zero marginal cost: picking the best from countless prompts, models and metrics (selection), and dissolving the evaluation target into the training distribution (leakage).

These can be gathered into a single expression. In addition to the bare denominator $\gamma=1/k_{\mathrm{eff}}$ of Section~\ref{sec:phantom}, let $m_{\mathrm{eff}}\ (\ge1)$ be the effective number of independent analyses, $\lambda\in[0,1]$ the degree of leakage, and $r$ the probability that a leaked path yields a positive even when the target is absent. We consider the case $r>\gamma$, in which leakage pushes the denominator up more strongly than the bare concentration. Under an independence-and-leakage approximation the effective denominator is
\begin{equation}
\boxed{\;\PrNC \approx 1-(1-\gamma_\lambda)^{m_{\mathrm{eff}}},\qquad \gamma_\lambda=(1-\lambda)\,\gamma+\lambda\,r\;}
\end{equation}
which reduces to the bare $\PrNC=\gamma=1/k_{\mathrm{eff}}$ at $m_{\mathrm{eff}}=1,\ \lambda=0$. The point is that the cap $\PrNC\le1$ enters automatically. The more analysis paths are stacked, the more the denominator saturates toward $1$ and the effective evidence $\LR_{\mathrm{actual}}=\PrC/\PrNC$ collapses toward $\PrC$. With effective options $k_{\mathrm{eff}}$, trying only about $0.69\,k_{\mathrm{eff}}$ analyses brings $\PrNC$ to $1/2$, beyond which a positive is obtained more often than not even when the target is absent.

Phantom evidence follows directly from this. The observer still assumes the denominator to be $1/N$, so the perceived likelihood ratio is $\LR_{\mathrm{perceived}}\approx qN$, where $q=\PrC$ is the probability that a genuine artifact looks convincing, corresponding to power. The gap between believed and actual evidence is then
\begin{equation}
\boxed{\;\Delta_{\mathrm{total}}=\log\frac{\LR_{\mathrm{perceived}}}{\LR_{\mathrm{actual}}}=\log\!\big(N\cdot\PrNC\big)\;}
\end{equation}
into which the denominator above may be substituted directly. When $m_{\mathrm{eff}}\gamma_\lambda\ll1$ this decomposes additively into three contributions:
\begin{equation}
\Delta_{\mathrm{total}}\;\approx\;
\underbrace{\log\frac{N}{k_{\mathrm{eff}}}}_{\substack{\text{nominal mistaken}\\\text{for effective}}}
\;+\;
\underbrace{\log m_{\mathrm{eff}}}_{\text{selection}}
\;+\;
\underbrace{\log\frac{\gamma_\lambda}{\gamma}}_{\text{leakage}}.
\end{equation}
\begin{itemize}
  \item Nominal mistaken for effective, $\log(N/k_{\mathrm{eff}})$: the gap between the vast space $N$ the observer assumes and the effective options $k_{\mathrm{eff}}$ the generator actually spans. Output dimension collapse produces it.
  \item Selection, $\log m_{\mathrm{eff}}$: the overestimate incurred by picking the best from effectively $m_{\mathrm{eff}}$ analysis paths. It is a logarithmic correction isomorphic to a multiple-comparison (Bonferroni) correction for choosing one item out of many, though not the same device (Section~\ref{sec:objections}); for the extreme-value derivation and the numbers, see Appendix~\ref{sec:extreme}.
  \item Leakage, $\log(\gamma_\lambda/\gamma)$: the amount by which the bare denominator $\gamma$ is contaminated to $\gamma_\lambda$ as evaluation data enters selection, normalization or training.
\end{itemize}
Intuitively, the observer has paid neither the description length of ``which mode the output is in'' ($\log N$ or $\log k_{\mathrm{eff}}$) nor that of ``which analysis path was chosen'' ($\log m_{\mathrm{eff}}$). Phantom evidence is the sum of this unpaid code length. (As the denominator approaches $1$ the three terms cannot be added independently, and the additive expression holds only as an upper bound; strictly, the boxed expression above is always the primary one. The running example already sits outside the small-signal regime, with $m_{\mathrm{eff}}\gamma_\lambda=0.4$, so its additive value $18.6$ bits exceeds the exact $18.3$ bits.) Figure~\ref{fig:delta} shows how $\Delta_{\mathrm{total}}$ swells with respect to the two principal terms, the nominal-for-effective confusion $\log(N/k_{\mathrm{eff}})$ and selection $\log m_{\mathrm{eff}}$. This probabilistic skeleton depends on neither field nor method. Only the substance of $m_{\mathrm{eff}}$ and of the bare denominator $\gamma$ changes: in quantitative research they become subsets of the sample and multiple testing; in qualitative research the choice of historical sources and post hoc hypothesis selection (HARKing). The same inflation by selection operates in both.

Which route an operation acts through determines which antidote works. Cherry-picking, sequential peeking and HARKing act through selection, and preregistration ($m_{\mathrm{eff}}\to1$) together with multiple-comparison correction works. Double dipping and circular analysis act mainly through leakage, are not repaired by correction, and fall only under independent verification ($\lambda\to0$). Overfitting is mixed, and a properly conducted held-out or out-of-distribution evaluation lowers both. Leakage nonetheless persists if preprocessing is done on all the data, and $m_{\mathrm{eff}}$ revives if the test set is used repeatedly~\autocite{blum2015}. When a fresh test set was built for a much-reused benchmark, however, the drop in accuracy was itself large, but its main cause was attributed to distribution shift rather than to adaptive overfitting from reuse~\autocite{recht2019}. The terminus of every route is the same: $\LR\to q\le1$, the disappearance of positive evidence (strictly $\log\LR\le0$, so observing $C$ slightly disconfirms $T$).

\begin{figure}[t]
\centering
\input{figs/fig_delta_en.tikz}
\caption{The three-term decomposition of phantom evidence (running example, in bits $=\log_2$). From the observer's perceived evidence $\log_2\Lambda_{\mathrm{perceived}}=19.8$, the confusion of nominal with effective, $\log(N/k_{\mathrm{eff}})$, subtracts $14.3$ and selection $\log m_{\mathrm{eff}}$ subtracts $4.1$, leaving the effective evidence $\log_2\Lambda_{\mathrm{actual}}=1.4$. The total of the drops is the phantom evidence, $\Delta_{\mathrm{total}}=18.3$ bits. The confusion term dominates; selection is less than a third of it. The dashed outline on the selection bar is the additive approximation $\log m_{\mathrm{eff}}=4.3$, and the smaller actual drop of $4.1$ is subadditivity (the running example, at $m_{\mathrm{eff}}\gamma_\lambda=0.4$, is already outside the small-signal regime). To convert to the bans of the main text, divide by $\log_2 10\approx3.32$.}
\label{fig:delta}
\end{figure}

\section{Three Layers and the Posterior: A Generative-AI Version of Ioannidis's PPV Argument}\label{sec:ppv}
So far we have argued at the prediction layer: whether an individual output derives from the target. The same $\LR$ structure extends isomorphically to at least three layers by changing the carrier of the attribute (Table~\ref{tab:layers}): the target fidelity of an individual prediction, the validity of a method, and the truth and reproducibility of a claim. We keep the base at the prediction and lift to the higher layers as needed.

\begin{table}[t]
\centering\small
\caption{The three layers over which $C$/$T$ and $\LR$ extend isomorphically. The base is placed at the prediction (fidelity) and lifted, as needed, to the method (validity) and the claim (truth). The reference set of $\pi=\Pr(T)$ changes with the layer.}
\label{tab:layers}
\begin{tabular}{@{}p{0.20\linewidth}p{0.36\linewidth}p{0.34\linewidth}@{}}
\toprule
Layer (carrier) & Meaning of $T$ & Apt term and lineage \\
\midrule
(1) Prediction or output (a single exhibit), the base & this output is not spurious but carries information derived from the target & target fidelity; machine-learning and AI-style prediction evaluation \\
(2) Method or model (the measurement and analysis pipeline) & the method captures the target truly, not via confounds, priors, or shortcuts & validity; method evaluation \\
(3) Claim or paper & the finding is true and reproduces & truth and reproducibility; Ioannidis's positive predictive value (PPV), the reproducibility crisis \\
\bottomrule
\end{tabular}
\end{table}

Carried through to the posterior probability, our diagnosis becomes a generative-AI version of the positive predictive value (PPV) argument of \textcite{ioannidis2005}. Let $\pi=\Pr(T)$ be the base rate of target fidelity: the probability, before the judgment $C$ is seen, that a given prediction truly reflects the target rather than being spurious. With prior odds $\rho=\pi/(1-\pi)$, the posterior is the prior multiplied by $\LR$:
\begin{equation}
\mathrm{PPV}=\Pr(T\mid C)=\frac{\LR\,\rho}{\LR\,\rho+1}.
\end{equation}
When the denominator swells and $\LR\to1$, we get $\mathrm{PPV}\to\pi$. The observation adds nothing and the posterior returns to the base rate. Now read the same expression at the claim layer (Table~\ref{tab:layers}, row 3). There $\pi$ is the fraction of hypotheses entertained in a field that are true, and $\mathrm{PPV}$ is the fraction of findings published as convincing results that are true. When \textcite{ioannidis2005} argued that most published research findings are false, the claim was that this $\mathrm{PPV}$ is low. Once looking convincing ceases to function as evidence ($\LR\to1$), passing through publication says nothing about whether a finding is true, and the fraction of published findings that are true stays at the base rate $\pi$ of the hypothesis pool. The reproducibility crisis is one principal route by which denominator collapse appears at the level of claims (low prior odds $\rho$ also contribute independently; see Section~\ref{sec:objections}).

The gap in log odds between the believed PPV and the true PPV is phantom evidence. What must be stressed first is that the likelihood ratio itself is not in error: $\LR$ is correct irrespective of when the hypothesis was formed (the likelihood principle). The harm lies in mistaking this one quantity for a more ambitious one: for the credibility of a selected hypothesis (a posterior probability), or for the success of a risky prediction in the sense of \textcite{meehl1978}. We do not commit to the likelihood principle here, but it is worth noting that the same conclusion follows even if one adopts it. What the principle protects is that, once the actually observed event and the two hypotheses being compared are fixed, the evidence depends on the likelihood function alone. That selection erodes evidence is not an exception to the principle but a consequence of the two items having been fixed wrongly: the observed event is not ``a prespecified single path was convincing'' but ``one of $m_{\mathrm{eff}}$ paths was convincing,'' and what is being compared is not the winner but the group of hypotheses. The gap between $\LR_{\mathrm{perceived}}$ and $\LR_{\mathrm{actual}}$ is not an error in the likelihood ratio but an error about what the likelihood ratio was computed for.

\begin{figure}[t]
\centering
\input{figs/fig_ppvladder_en.tikz}
\caption{The ladder of evidence, the two ceilings, and the two directions of the prescription (running example, in bits). Three ladders rise from the same footing (the prior $\pi$). Mistaking the denominator for $1/N$, the observer climbs the perceived ladder (red, $19.8$). What can actually be climbed is the rung set by the generator's output concentration alone ($5.5$), and including the selection $m_{\mathrm{eff}}=20$ of the running example it falls further to $1.4$. The total gap $\Delta_{\mathrm{total}}=18.3$ is the overestimate of the hit rate (matching the decomposition of Figure~\ref{fig:delta}). The dashed lines are the two ceilings; that the actual ladder sits exactly at the ceiling on evidence, $\log_2 k_{\mathrm{eff}}=5.6$, shows that the bound $\log\LR_{\mathrm{actual}}\le\log k_{\mathrm{eff}}$ is the binding constraint. The two arrows on the right are the two directions of the prescription, and they move different things: direction 1 pushes the blue ceiling up, direction 2 pulls the red ladder down to the measured denominator. Neither depends on $\pi$ or $q$.}
\label{fig:ppvladder}
\end{figure}
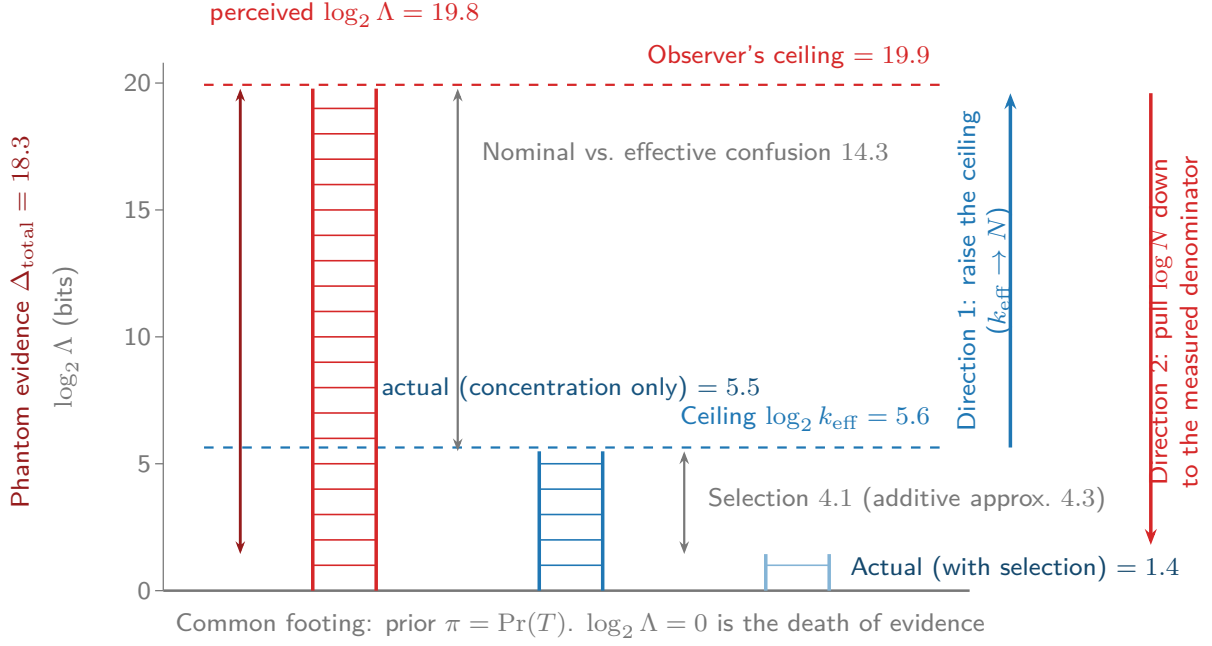

What is decisive is the gap between the hit rate the observer believes and the actual hit rate. The observer mistakes the denominator for $1/N$, so uses $\LR_{\mathrm{perceived}}\approx qN$ with $q=\PrC$ and, across nearly the whole range, believes $\mathrm{PPV}_{\mathrm{perceived}}\approx1$, near certainty. But the true $\LR_{\mathrm{actual}}=q/\PrNC$ collapses toward $1$ as the denominator $\PrNC$ approaches the numerator $q$. Measuring the gap in log odds, the prior $\rho$ and the power $q$ cancel:
\begin{equation}
\boxed{\;\operatorname{logit}\mathrm{PPV}_{\mathrm{perceived}}-\operatorname{logit}\mathrm{PPV}_{\mathrm{actual}}
=\log\frac{\LR_{\mathrm{perceived}}}{\LR_{\mathrm{actual}}}=\log\big(N\,\PrNC\big)=\Delta_{\mathrm{total}}\;}
\end{equation}
so the overestimate of the hit rate in log odds is exactly the phantom evidence $\Delta_{\mathrm{total}}$ of Section~\ref{sec:mechanisms}, independent of $\pi$ and $q$. Only three things inflate the gap: $N/k_{\mathrm{eff}}$ (nominal mistaken for effective), $m_{\mathrm{eff}}$ (selection) and leakage. Note also that $1-\mathrm{PPV}_{\mathrm{perceived}}\propto1/N$: the larger the assumed space, the more one feels it cannot be wrong. This is the information-theoretic counterpart of the transposed conditional, mistaking ``$\PrNC$ is small'' for ``$\Pr(\neg T\mid C)$ is small,'' the prosecutor's fallacy, together with the base-rate neglect behind it; and generative AI is a device that mass-produces the confusion (Figure~\ref{fig:ppvladder}).

From this single identity one can read off what has to be moved for the overestimate to change (Table~\ref{tab:compstat}). What is not obvious is that the usual proxies for research quality, a high prior and high power, do not affect the size of the overestimate, and that ``bigger and higher-resolution'' works in the wrong direction. $\Delta_{\mathrm{total}}$ also falls if $m_{\mathrm{eff}}$ or $\lambda$ is lowered, but only one quantity reduces it by being increased. That quantity is the effective number of options $k_{\mathrm{eff}}$, which shrinks the overestimate and at the same time raises the ceiling $\log\LR_{\mathrm{actual}}\le\log k_{\mathrm{eff}}$ on the effective evidence.

\begin{table}[t]
\centering\small
\caption{Comparative statics of phantom evidence $\Delta_{\mathrm{total}}=\log(N\,\PrNC)$: how the overestimate of the hit rate moves when each quantity is increased.}
\label{tab:compstat}
\begin{tabular}{@{}p{0.14\linewidth}p{0.28\linewidth}p{0.48\linewidth}@{}}
\toprule
Direction & Quantity & Meaning \\
\midrule
Increases it & nominal $N$, selection $m_{\mathrm{eff}}$, leakage $\lambda$ & the more resolution and candidates are raised, trials repeated, and evaluation mixed into training, the larger the overestimate \\
No effect on $\Delta_{\mathrm{total}}$ & prior $\pi$, power $q$ & the overestimate is not a problem of a lenient prior or of low power\textsuperscript{a} \\
Decreases it & effective options $k_{\mathrm{eff}}$ & the only quantity that shrinks the overestimate and simultaneously raises the ceiling on the effective evidence (Section~\ref{sec:prescription}, direction 1)\textsuperscript{b} \\
\bottomrule
\end{tabular}

\smallskip
{\footnotesize\raggedright \textsuperscript{a}What cancels is the difference in log odds. The difference on the probability scale, and the level of the true PPV itself, do depend on $\pi$ and $q$.\\
\textsuperscript{b}At $\lambda=1$ we have $\gamma_\lambda=r$, and $k_{\mathrm{eff}}$ ceases to act.\par}
\end{table}

What do the two directions mean in PPV terms? The two directions of the prescription (Section~\ref{sec:prescription}) act on different parts of Figure~\ref{fig:ppvladder}. Direction 1, raising $k_{\mathrm{eff}}$ toward $N$, lifts the ceiling on $\LR_{\mathrm{actual}}$ and so raises the true PPV; direction 2, measuring the denominator with negative controls, pulls $\LR_{\mathrm{perceived}}$ down and corrects the believed PPV to the true one, without raising the PPV itself. What each direction involves is set out in Section~\ref{sec:prescription}.

\section{Worked Example: Reconstructing Visual Images from Brain Activity}\label{sec:sites}
The mechanisms above are not confined to thought experiments. Here we take the reconstruction of visual images from brain activity as a worked example and follow it down to numbers, in order to see how operations that inflate the denominator and evaluations that correctly control it come apart on real data. (The low replication rates in psychology~\autocite{osc2015}, the troubling trends in machine learning~\autocite{lipton2019} and the leakage-driven reproducibility crisis~\autocite{kapoor2023} can all be read isomorphically once lifted to the claim layer; see Section~\ref{sec:ppv}.)

Consider the reconstruction of visual experience from brain activity~\autocite{kamitani2025}. Here the target is not the stimulus image itself but the latent representation, or latent features, of the visual image encoded by brain activity: there is no image inside the brain, only a representation of the visual information in some other format. The target features must, however, be specified in a feature space that is fixed on the stimulus side, independently of the decoder and the generator. If the target is defined in the space that the decoder or generator optimizes, the evaluation is circular by construction (see the discussion of circular metrics below). $T$ is ``the reconstruction carries information derived from this target,'' that is, the content of the output is attributable to, and carries information about, the latent features encoded by the very brain activity in question, rather than a prior distribution or a category. $C$ is ``the reconstructed image is judged to look plausible, or real.'' The danger lies in the fact that the two can come apart.

\textcite{shirakawa2025} showed that, in recent reconstruction methods using large datasets and text-guided diffusion models, the apparent realism stems chiefly from two sources, the diffusion model's generative prior (its power to produce plausible natural images) and classification into trained categories, and does not reflect genuine reconstruction. Behind this lies a circumstance on the decoder side. Because the (semantic) features of the training data are clustered and unevenly distributed, the decoder's predictions shrink into the low-dimensional subspace spanned by the training features. This is output dimension collapse, a term introduced by \textcite{shirakawa2025}; \textcite{otsuka2025} give its mathematical analysis and a remedy through sparsity. Moreover, the semantic clusters of the training and evaluation sets overlap, so the evaluation is not a zero-shot test and the collapse does not show up in the scores. When these combine, the generative prior fills in the details and a photorealistic image appears, that is, $C$ even under $\neg T$, even though the brain signal's contribution reaches no further than the category level. $\PrNC$ is high.

Furthermore, \textcite{shirakawa2025reanalysis} reanalyzed a published reconstruction study, \textcite{koidemajima2024}, and showed that two denominator-inflating operations are at work on real data: selective reporting of the best-performing examples through comparisons at multiple levels (selection), and circular metrics that do not reflect perceptual accuracy (leakage). Under a fair baseline comparison the study's key innovations showed no discernible advantage ($\LR\approx1$). Two disclosures are due here. First, that reanalysis is a preprint under review, not a refereed result. Second, its authors include the present authors: it is a reanalysis by a group other than the original authors, but not by a group independent of us. Denominator inflation is nevertheless not an abstract worry.

Working through the numbers exposes the crux. If the criterion is placed at ``does it look realistic?'', genuine and spurious reconstructions look equally realistic, since the diffusion prior guarantees it. Suppose $\PrC\approx 0.95$ and $\PrNC\approx 0.90$: then $\LR\approx 1.06$, and no more: realism carries almost no evidence. But replace the criterion with ``can it win a forced-choice identification against many distractors?'', that is, can the true stimulus be picked out of $N=100$ candidates, and the picture changes entirely. A genuine reconstruction yields a high accuracy ($\PrC\approx 0.9$), while a spurious one, if the identification task is hard enough, approaches chance ($\PrNC\approx 1/N=0.01$). Now $\LR\approx 90$ and $\log\LR$ jumps from essentially $0$ to just under $2$ (about $1.95$ bans). For the very same artifact, the evidential value changes by two orders of magnitude depending on how the evaluation is designed.

Identification accuracy is not, however, an unconditionally safe metric. As Shirakawa et al.\ themselves show, when the evaluation metric shares the same feature space as the reconstruction model, identification accuracy is inflated circularly: reconstructing from decoded CLIP features alone yields images that differ substantially from the targets, yet pairwise identification accuracy in that CLIP feature space remains high at about $75\%$ (chance $50\%$), while an independent measure (pixel correlation) drops to near chance~\autocite{shirakawa2025reanalysis}. Since the reconstruction is built by optimizing the decoded CLIP features, identification measured in the same space merely confirms the algorithm's internal consistency and does not measure whether the image resembles the target.

This is the form that the phantom evidence of Section~\ref{sec:phantom} and the three-term decomposition of Section~\ref{sec:mechanisms} take on numbers in this worked example: the $75\%$ under a circular metric corresponds directly to leakage $\lambda$, and the reporting of best examples across multiple levels to selection $m_{\mathrm{eff}}$. Raising the resolution and dimensionality of the output does not increase the collapsed effective dimension, and since the added degrees of freedom are merely filled in for free by the generative prior, what grows is only phantom evidence.

To drive $\PrNC$ down to chance, that is, to bring $k_{\mathrm{eff}}$ close to $N$, requires a hard discrimination (for the design, see Section~\ref{sec:prescription}(4)). A two-alternative pairwise identification, by contrast, in which one asks which of the true stimulus and a single distractor the output corresponds to, has a per-trial denominator as large as $1/2$. Repeating trials lowers it, but the gain saturates with the correlation between trials (Appendix~\ref{sec:subadditivity}). The more fundamental weakness lies in the design of the distractors: if they come from a different category, the test can be won on a single coarse dimension such as brightness or category alone, and an output that carries none of the detail of the visual image can still score high. Adopting the form of an identification test does not by itself lower the denominator. What is rejected is not reconstruction itself but evaluation that treats realism as evidence: a practice that scores a quantity which can be high even under $\neg T$ and never measures discrimination information. A genuine reconstruction retains a high $\LR$ under correct discriminative evaluation. The metric must be the accuracy of discrimination rather than the quality of generation, namely realism, because that is the only design that explicitly holds $\PrNC$ low.

\section{Prescription: Raise the Ceiling and Measure the Denominator}\label{sec:prescription}
This framework demands not epistemic pessimism but a concrete methodological reorganization. The prescriptions below rest on one normative premise: that a published claim should be priced by the amount of evidence it gives the community. This premise does not deny the value of exploratory research. An output with $\LR\approx1$ can be extremely useful as a hypothesis generator or a tool (Section~\ref{sec:scope}). What is demanded is not a ban on exploration but a refusal to price the products of exploration as products of confirmation, that is, an explicit statement of which currency one is reporting in. Phantom evidence $\Delta_{\mathrm{total}}=\log(N\,\PrNC)$ can be driven toward $0$ from either end, by raising the effective $k_{\mathrm{eff}}$ toward the nominal $N$, or by replacing the observer's assumed denominator $1/N$ with a measured $\PrNC$. Two complementary directions follow.

\subsection*{Direction 1: Make Outputs Genuinely Diverse ($k_{\mathrm{eff}}\to N$)}
Widen the range of outputs the system can actually reach, that is, its true diversity and coverage. Restoring effective dimensions that have collapsed brings $k_{\mathrm{eff}}$ closer to the nominal $N$, and the ceiling on the effective evidence, $\log\LR_{\mathrm{actual}}\le\log k_{\mathrm{eff}}$, rises toward $\log N$. What is being lifted is the bound $\log\LR_{\mathrm{actual}}\le\log k_{\mathrm{eff}}$ and nothing else; no amount of work on appearance substitutes for it (Section~\ref{sec:lr}). This is the direction of capability, of building good science, and it lifts $\LR_{\mathrm{actual}}$ from below toward $\LR_{\mathrm{perceived}}$.

The concrete lever is to decompose the output into independently specifiable parts, that is, a factorized (compositional, modular) design. If the output is composed of $J$ independently specifiable modules, each discriminable into $c$ states, the number of distinguishable outputs grows combinatorially in principle as $k_{\mathrm{eff}}\sim c^{J}$. This is a domain-neutral claim about the structure of the output space, and its implementation differs by field: in an evaluation of model capability it amounts to requiring a system to get several independently scored aspects right at once, rather than a single aggregate score. In visual reconstruction, factorized (sparse) feature encoding is one such implementation, and has been shown to enable zero-shot recovery with small datasets~\autocite{otsuka2025}. This exponential law, however, requires six conditions. (i) The reachable set really has a product structure. (ii) Each module is discriminable into $c$ levels above the noise. (iii) The decoding errors of different modules are uncorrelated; if they are correlated, the count shrinks to $c^{J_{\mathrm{eff}}}$, isomorphically to Appendix~\ref{sec:subadditivity}. (iv) The judgment $C$ is sensitive to all modules. (v) Within each module, the $\neg T$ output is close to uniform over the $c$ states. (vi) Under $\neg T$, the outputs of different modules are independent.

Condition (vi) is easily violated: a generative prior fixes colour, shape and context together once the category is fixed, so several modules hit at once even under $\neg T$, giving $\gamma>\prod_j\gamma_j$ and shrinking the effective count to $c^{J_{\mathrm{eff}}}$. And if each factor is itself concentrated, then even $\gamma=\prod_j\gamma_j$ far exceeds $c^{-J}$. Condition (iv) means that directions 1 and 2 are not independent: since $k_{\mathrm{eff}}=1/\gamma$ is a quantity relative to the judgment rule, a coarse judgment leaves $k_{\mathrm{eff}}$ unchanged however much compositionality is increased. By contrast, in a low-effective-rank code that has collapsed onto a few prototypes (output dimension collapse, Section~\ref{sec:sites}), raising the nominal $N$ exponentially leaves $k_{\mathrm{eff}}$ orders of magnitude smaller. The cost is high. One must stop outsourcing the work to the prior and actually push target-specific information through to the output.

One caveat governs this whole direction. Widening coverage is not the same thing as raising $k_{\mathrm{eff}}$: if the added output states are not discriminable above the noise, the second condition fails and $k_{\mathrm{eff}}$ does not move. And because $\LR_{\mathrm{actual}}=\PrC\cdot k_{\mathrm{eff}}$, coverage bought at the cost of $\PrC$ lifts the ceiling without lifting the evidence. Direction 1 must therefore be reported as two numbers, coverage together with per-target accuracy, and never as coverage alone (Section~\ref{sec:scope}).

\subsection*{Direction 2: Measure the Denominator (Replace $\log N$ with a Measured Value)}
The second direction does not increase the effective evidence but exposes phantom evidence by replacing the observer's mistaken denominator $1/N$ with the measured $\PrNC$. This is the direction of institutionalized integrity, of bringing things to light; it is cheap and always advisable. The integrity meant here is not a personal virtue. Failing to measure the denominator is usually not misconduct but a consequence of convention and of the reporting format (the garden of forking paths~\autocite{gelman2014garden} is walked unintentionally). What is required is a procedure that makes measurement the default. The central principle is to invest in negative controls rather than in positive examples, which is nothing other than the table of absence and exclusion that Bacon set out four centuries ago, the operation of checking that a property fails to appear where it should not. The five items below fall into two groups. The first three, (1) to (3), measure the denominator itself, and correspond one-to-one to the three quantities that swell $\PrNC$ (the bare denominator $\gamma$, selection $m_{\mathrm{eff}}$, and leakage $\lambda$; Section~\ref{sec:mechanisms}). The last two, (4) and (5), carry the measured denominator into the institutions of evaluation, reporting and verification.

\smallskip\noindent\textbf{A. Measure the denominator ((1) to (3))}\smallskip

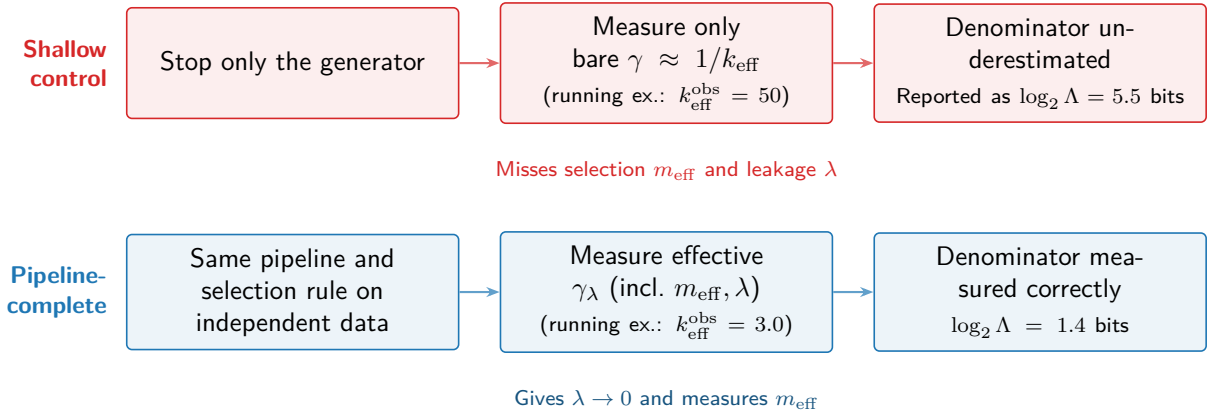
\begin{figure}[t]
\centering
\input{figs/fig_negcontrol_en.tikz}
\caption{Shallow against whole (self-contained) negative control. Top: a shallow check that halts only the generator and measures the bare $\gamma=1/k_{\mathrm{eff}}$ misses selection $m_{\mathrm{eff}}$ and leakage $\lambda$, and so underestimates the denominator. Bottom: a self-contained control that runs the same analysis procedure and the same selection rule in full on independent data measures the selection-inclusive effective denominator $\PrNC$, giving both $\lambda\to0$ and an empirical measurement of $m_{\mathrm{eff}}$.}
\label{fig:negcontrol}
\end{figure}

(1) Construct adversarial negative controls (measure the bare denominator $\gamma$). Build a $\neg T$ condition that matches everything driving $C$ except $T$: label-permuted inputs, ablated models, sham targets. A good negative control is a device that actually runs and checks the counterfactual ``if $T$ were false, $C$ should not appear,'' and evidence resides only in $C$ failing to appear under that control. But the control must be run through the whole analysis procedure, including the selection rules of (2) and the data splits of (3). A shallow control that merely switches off the generator sees only the bare $\gamma$ and underestimates the denominator (Figure~\ref{fig:negcontrol}).

Estimation itself is not hard: from a $k$-alternative forced choice with accuracy $p$, if the negative control truly falls to chance $\PrNC\approx1/k$, then $\log\LR=\log(p\,k)$ (in the worked example, $k=100$ and $p\approx0.9$ give $\LR\approx90$). If cues leak into the distractors then $\PrNC>1/k$ and the true value is smaller, so $\log(p\,k)$ should be read as an optimistic estimate. What should be reported is not the nominal number of options $k$ or the output resolution but the measured effective number of options $k_{\mathrm{eff}}^{\mathrm{obs}}=1/\PrNC$ that the negative control actually spans.

(2) Preregister and freeze the evaluation criteria (close off selection $m_{\mathrm{eff}}$). Fix the criteria for judging $C$ before seeing the outputs~\autocite{nosek2018}. If the criteria can be adjusted post hoc, the denominator is fitted after the fact and $\PrNC$ can no longer be measured. Preregistration is an institutional device that protects the completeness of the denominator ($m_{\mathrm{eff}}\to1$).

(3) Control contamination and leakage (close off leakage $\lambda$). Use held-out tasks from which overlap with the training data has been explicitly excluded. Leakage is the most common route by which reproducibility is destroyed, across fields~\autocite{kapoor2023}, and it quietly opens the path ``a correct answer appears even under $\neg T$.'' In evaluations of large language model ability this appears as the control of benchmark contamination. A performance report that does not control it says nothing about $\log\LR$ ($\lambda\to0$).

\smallskip\noindent\textbf{B. Carry the measured denominator into institutions ((4) and (5))}\smallskip

(4) Keep generation quality and evidence in separate slots. Ask not only ``is the output plausible?'' but also ``can the system reject a plausible fake?'' Generation quality is not worthless in itself: making visible what has been recovered, and how far, is in a task such as visual reconstruction the very content of the result, and it should be shown. But generation quality is a metric that looks only at the numerator; only discrimination that rejects fakes measures the denominator. Accuracy rates and galleries of generated artifacts therefore belong in the slot that describes what was obtained, while the slot read as an amount of evidence should hold discriminative performance alone (discrimination against negative controls and, where possible, $\log\LR$). The reporting format is an institutional force that dictates what researchers optimize, and a format that counts the numerator reproduces a culture that inflates only the numerator.

Two caveats apply on the discriminative side. First, discrimination accuracy is not unconditionally safe either: two-alternative pairwise identification is inflated by category and low-level structure alone (Section~\ref{sec:sites}), so draw distractors from within the same category, control low-level cues, and validate on semantic clusters unseen during training (zero-shot, within-category identification). Second, once a single metric becomes the currency of achievement, algorithms and research practice deform themselves to maximize it and shortcuts that raise it are rewarded~\autocite{geirhos2020shortcut}. The evidence slot should therefore not be collapsed into a single number but designed as several tests that can fail independently, and a reported $\log\LR$ must travel with the design of the negative controls and distractors used to estimate it. Identification against a set of candidates has been the standard discriminative test in this literature since \textcite{kay2008}, but being standard does not license making it the objective to be maximized.

The other pillar of reporting is disclosure. For direction 1, report coverage and per-target accuracy together, and disclose in addition the effective number of analysis paths tried, $m_{\mathrm{eff}}$, and the selection rule, since the community can only inspect what is reported. This is a layer of transparency that operates independently of the epistemic prescriptions, and its justification is taken up in Section~\ref{sec:objections}. An isomorphic requirement holds for disclosing the provenance of post hoc hypothesis selection (HARKing), an issue that extends even to research that uses no statistics at all.

(5) Design the oracle. By an oracle we mean an independent verifier that a fake cannot cheaply pass. Where (4) fixed what counts as evidence, the question here is who passes that judgment, and how. Ultimately, convert ``looks convincing'' into ``passes an oracle.'' Since these artifacts are now produced at industrial scale, individual vigilance cannot keep pace with them. What is needed is a device whose very passing structurally holds $\PrNC$ low. There are five levers.
\begin{itemize}\setlength{\itemsep}{1pt}
\item[(a)] Soundness: make it a verifier whose acceptance logically entails the target (a machine-checkable proof is the typical case). Within that range, $\Pr(C\mid\neg T_{\mathrm{formal}})=0$ by construction for the formalized target $T_{\mathrm{formal}}$. But the validity of the specification, the implementation soundness of the checker, and unproven axioms all lie outside soundness, so the measured denominator is not $0$.
\item[(b)] Out-of-loop freshness: use data after the training cutoff, or a secret or rotating oracle, making it a target the generator has not optimized against.
\item[(c)] External grounding: make acceptance depend not merely on the appearance of the output but on physics, causation or fresh observations the generator cannot control (a physical experiment under preregistered conditions, for instance).
\item[(d)] Demand exhibits: require an inspectable proof, a reproduction procedure or a preregistered prediction rather than a convincing appearance.
\item[(e)] Asymmetric cost of faking: design it so that actually achieving the target is the cheapest strategy for passing (costly signaling in the sense of \textcite{spence1973}).
\end{itemize}
Item (a) is strongest and (b) to (e) complement it. The essential point is single: keep the oracle outside the optimization loop (Section~\ref{sec:objections}). LLM-as-judge~\autocite{zheng2023}, which lets a generative system do the grading, is a device in which the oracle itself inflates $\PrNC$ whenever the judge shares the generator's prior and blind spots and sees only the appearance of the output (evaluators have been shown to recognize and prefer their own generations~\autocite{panickssery2024}). Give the judge reference answers or external information the generator cannot control, use a different model family, and keep it outside the loop, and it can be a useful screen. But so long as the verdict rests on the appearance of the output, it is not an oracle with the soundness of (a).

Three pitfalls of negative controls. First, a clean control cannot always be built: when the surface features used in judging are causally downstream of $T$, an operation that removes only $T$ while preserving the surface is hard to construct, and one must fall back on bounding the residual $\PrNC$ from the agreement of several imperfect controls. Second, the control or the identification task itself inflates the denominator if badly designed (if low-level cues from a generative prior leak into the distractors or the metric, $C$ appears even under $\neg T$)~\autocite{lapuschkin2019}. Third, a denominator has an address. For the same method it changes with the level at which it is measured: an i.i.d.\ held-out evaluation, structured cross-validation with subjects or sessions as the unit, or a genuine distribution shift to unseen data (zero-shot), which is the most demanding and puts the numerator $\PrC$ to the test as well. None of these proves external validity: any new distribution is itself another closed world, and what one obtains is not a guarantee but a more demanding opportunity for refutation. What should be reported is therefore not only the value of the denominator but which $\neg T$ it is a denominator for, and at which level it was measured.

Both are required. The negative controls of direction 2 tell us where we stand, that $\LR$ is in fact small, but they do not make $\LR$ large. To increase evidence genuinely one must widen coverage in direction 1 and lift the ceiling. In Bacon's vocabulary, direction 2 breaks the anticipations of nature with the table of absence and the crucial instance (\textit{instantia crucis}). Direction 1 has no single Baconian counterpart. The nearest is the diversity of instances that Bacon required as the precondition for exclusion, in what he called the interpretation of nature (\textit{interpretatio naturae}). Reliable science is the enterprise of turning these two continually.

\section{Scope: Where the Framework Holds and Where It Weakens}\label{sec:scope}
No formalization deserves trust until its range of application is made explicit. This section states in turn the regions where the framework holds strongly, where it weakens, and the limits of the framework itself.

Where it holds strongly. The framework holds strongly wherever a powerful generator can produce $C$ without $T$: where the output space is high-dimensional and open, judgment is subjective and plausibility-dependent, and the criterion ``convincing'' is cheap to satisfy. Free-form generation, open-ended hypothesis generation and evaluation indexed to human impression are the clearest cases. Here denominator collapse dominates and our warning applies most sharply.

Where it weakens. In regions where independent verification is cheap, $\PrNC$ is intrinsically low. Passing a formal-proof checker, predicting held-out fresh data, passing a physical experiment with pre-committed criteria: here satisfying $C$ itself cannot be achieved cheaply by a fake. The prescription is therefore the oracle of Section~\ref{sec:prescription}(5).

Where numerator saturation fails. There are settings in which the numerator saturation $\PrC\approx1$ assumed in Section~\ref{sec:lr} does not hold. Psychology and neuroscience are known for chronic low power, and the demanding discriminative evaluation this paper itself recommends lowers the numerator $\PrC$ even as it lowers the denominator $\PrNC$ (the sensitivity-specificity trade-off). The same trade-off constrains direction 1 from the other side, which is why coverage must be reported together with per-target accuracy (Section~\ref{sec:prescription}). In these regions the numerator too governs evidential value. The correct prescription is therefore not ``lower the denominator alone'' but ``maximize $\log\LR$, choosing designs that lower the denominator while preserving the numerator as far as possible.'' The slogan ``evidence lives in the denominator'' is, in this sense, strictly an approximate statement about the saturated regime.

Our analysis is static. Extended to the time axis, phantom evidence turns out to have a lifetime, and generative AI speeds up only the clock that manufactures it (Appendix~\ref{sec:lifetime}).

Limits of the framework itself. First, this is Bayesian inference itself and in that sense mathematically not new. Our contribution is to locate the generative-AI threat precisely in the denominator of inference (phantom evidence) and to derive a prescriptive reorientation of evaluation culture. Second, $\LR$ concerns evidential value, not utility. Generative AI can be extremely useful as a tool even where $\LR\approx1$ for truth claims. Third, there are scientific domains in which negative controls are difficult to construct in principle; the honest conclusion there is not pessimism but humility.

Fourth, the three-term decomposition (Section~\ref{sec:mechanisms}) is not a formula for an effect size in an open space: unless the set of analysis paths that could have been tried is uniquely determined, $m_{\mathrm{eff}}$ is not unique either. What the decomposition provides is a map of the existence of phantom evidence and of the routes by which it enters, not a point estimate of $\Delta$ (for reporting in practice see Section~\ref{sec:prescription}(4)).

Fifth, $\Delta_{\mathrm{total}}=\log(N\,\PrNC)$ is relative to two specifications, the nominal space $N$ the observer takes as default and the ensemble into which $\neg T$ is made concrete (Section~\ref{sec:objections}), so $\Delta_{\mathrm{total}}$ can be reported only as a difference against an explicitly stated baseline, never as an absolute quantity. Since $\Delta_{\mathrm{total}}$ is defined as the logarithm of a ratio of likelihood ratios, moreover, the share of the selection penalty that enters through prior dilution lies in principle outside $\Delta_{\mathrm{total}}$; our decomposition is a ledger that assigns the whole selection term to the denominator side.

Sixth, the core of direction 1, that factorized encoding pushes $k_{\mathrm{eff}}$ up combinatorially, is an empirical bet about representation learning and not a theorem. This asymmetry, direction 2 being cheap and certain while direction 1 is expensive and uncertain, is also why we ask for direction 2 first. Finally, let us apply the paper to itself: the worked example (Section~\ref{sec:sites}) derives from the authors' own prior work and was selected from among other candidate cases.

\section{Open Questions: Points Raised Against the Framework}\label{sec:objections}
The contours of a framework show most clearly in how it answers the questions put to it. We take up the main ones here.

``AI detectors will spot the fakes.'' Detection is a matter of provenance, not of truth. Even if an output can be judged AI-derived, whether it is true is a separate matter, and conversely detectors are powerless against convincing fakes made by humans. Detection also falls into an unstable arms race, and paraphrasing attacks degrade current detectors substantially~\autocite{sadasivan2023}. Our claim is not about AI detection but about the structure of the divergence between persuasiveness and truth.

``As models improve, $C$ will come to track $T$.'' That holds only when the training objective ties $C$ to $T$, that is, for the sound oracle or the out-of-loop oracle of Section~\ref{sec:prescription}. Place a learned discriminator, a tunable metric or a negative control inside the training loop as the optimization target, and the generator learns only to appear to pass that oracle, exactly as the generator of a generative adversarial network (GAN) learns to defeat its discriminator~\autocite{goodfellow2020}. In a GAN, under the standard setting that mixes real and generated data with equal probability, the odds of the optimal discriminator against a fixed generator equal exactly the likelihood ratio $p_{\mathrm{data}}/p_g$. At the global optimum, where the generator matches the distribution, they collapse to $1$---dynamics isomorphic to our $\LR\to1$. (This correspondence applies directly to a provenance-based oracle, one that asks whether an output is real-data-derived. Its generalization to truth-based oracles holds through the condition that the oracle be a function of the appearance of the output alone.) Hence, in the ideal limit where the acceptance decision is a function only of outputs under the generator's control and the oracle is built into the optimization loop, $C$ loses discriminative power and $\LR\to1$ (Goodhart's law). This is not an absolute impossibility but an asymptotic degradation of evidential power. A learned oracle that sees only the appearance of the output must not be placed in the same loop as the generator.

``This is just Bayes; there is nothing new.'' The mathematical skeleton is Bayesian. But the diagnostic value lies in which term of existing inference theory the new threat of generative AI is located in. The culture of frequentist significance testing has steered us toward counting positive rejections. The likelihood-ratio framework explicitly relocates evidential value to the denominator and names where the false positives of the generative-AI era arise.

More specifically, the selection term $\log m_{\mathrm{eff}}$ is isomorphic to a familiar Bayesian ledger. When the best of $m$ candidates, $H^\ast$, is selected and reported, there are only two correct ways to keep the books. Treating the group as a composite hypothesis, the Bayes factor is the marginal likelihood ratio, that is, the weighted average of the pointwise ratios under within-group prior weights $w_j$ (with $w_j\ge0,\ \sum_j w_j=1$), namely $\sum_j w_j\LR_j\le\max_j\LR_j$; or, looking at the winner individually, its prior is diluted to $w_{H^\ast}$. These are two ways of writing the same posterior odds, not two penalties to be charged twice. With $w_j=1/m$ and a dominant winner, either route gives a discount of $\log m$, matching our selection term. Thus the appearance is the winner's pointwise likelihood ratio while the reality is the group's marginal likelihood ratio, the same structure as our $\LR_{\mathrm{perceived}}$ against $\LR_{\mathrm{actual}}$. The device is not identical, however: the Bayesian discount enters the numerator (through marginalization and prior dilution), whereas our denominator version saturates at $\log k_{\mathrm{eff}}$. The Bayesian Occam's razor~\autocite{mackay2003} likewise arises from the same discipline while entering a different term: a flexible hypothesis spreads its predictive mass thin and thereby lowers the numerator, whereas in our version it is the smallness of $k_{\mathrm{eff}}$, the narrow concentration of the output, that raises the denominator. The mechanisms run in opposite directions, but one thing is common to both. Get right what is being assigned probability, and the discount enters without being inserted by hand. Our contribution is to name in which term, and by how much, generative AI produces that discount, and to connect it to a procedure for measuring it (negative controls).

``Surely the prior probability is what matters most.'' The prior odds do matter, but they are the subjective term, dependent on a field's background knowledge. What generative AI attacks is $\LR$, the part of inference relatively more transmissible across fields than the prior odds. $\LR$ is not fully objective either: the denominator $\PrNC$ is a conditioning on the composite hypothesis ``the negation of the target'' and depends on the make-up of $\neg T$. In practice $\neg T$ must be made concrete as identifiable rival hypotheses, such as leakage or a generative fake, and measured against them. Moreover, if generative AI floods the population with $\neg T$ artifacts, the base rate of $T$ in the population, and hence the prior odds, can also degrade. This does not weaken our claim: because the threat reaches the prior odds as well as $\LR$, it strengthens it.

``A competent evaluator never uses $1/N$; they measure chance with a permutation test.'' The point is well taken, and our warning is a generalization of exactly that practice. Measuring the null empirically rather than assuming a theoretical one is the established statistical response to large-scale selection~\autocite{efron2004}, and the empirical demonstration that an assumed null can be off by an order of magnitude is by now familiar in neuroimaging~\autocite{eklund2016}. Two remarks are in order. First, a null distribution built inside the same pipeline shares the generative prior, the selection rule and the leakage: the shallow negative control that halts only the generator and measures the bare $\gamma$ (Figure~\ref{fig:negcontrol}) is of this kind, and it misses $m_{\mathrm{eff}}$ and $\lambda$. Second, in this paper $N$ is the default baseline an audience implicitly uses, so $\Delta_{\mathrm{total}}$ measures not a property of the research itself but the gap that opens between reporting and reception. Reporting it therefore requires making explicit which denominator is being treated as the naive default.

``Is the cost of disclosure not larger?'' The requirement of the transparency layer (Section~\ref{sec:prescription}(4)) can be justified as weak dominance against the private cost of journal space: if selection does not affect the evidence, disclosing it costs almost nothing, and if it does, concealing it misleads the reader. Being strictly better in one case and no worse in the other, disclosure is an admissible option for anyone on either side of the dispute about whether selection erodes evidence. Two qualifications sit outside that accounting. First, above all, the adverse selection in which an author who honestly reports the number of paths fares worse in review than a competitor who does not; this has to be absorbed institutionally, by editorial policies that do not treat disclosure as a demerit. Second, $m_{\mathrm{eff}}$ cannot be counted uniquely (Section~\ref{sec:scope}). What should be required is therefore not a single number but an auditable record of the paths actually run together with a description of the selection rule, with any number read as a lower bound.

\section{Conclusion: The Return to the Negative Control}
Generative AI has not put a new question to science. It has taken an old error, namely marveling at the breadth of the nominal space while failing to measure the true denominator, and made that error repeat itself at incomparably greater scale, through the mass production of persuasiveness at near-zero marginal cost.

The likelihood ratio $\LR = \PrC / \PrNC$ makes this error a measurable quantity. Evidence lives not in the numerator, looking convincing, but in the denominator, the probability of looking convincing even when the target is absent, and phantom evidence, whose full ledger is $\Delta_{\mathrm{total}}=\log(N\,\PrNC)$, is the magnitude of the mistake. Generative AI cheaply inflates the denominator, collapsing $\LR$ toward $1$ wherever a generator's expressive power outruns any constraint that ties its output to the target. The prescription has two directions. Raise the effective $k_{\mathrm{eff}}$ toward the nominal $N$ by making outputs genuinely diverse (capability), and at the same time measure the true denominator with negative controls (integrity). Reliable science is not the enterprise of accumulating convincing artifacts. It is the enterprise of measuring the effective number of options that a negative control actually spans, $k_{\mathrm{eff}}^{\mathrm{obs}}$, and continuing to sort phantom evidence from actual evidence.

\medskip\noindent Historical coda. This diagnosis is not new. Four centuries ago, in the \textit{Novum Organum}~\autocite{bacon1620}, Francis Bacon warned against the anticipations of nature (\textit{Anticipationes Naturae}), a mode that generalizes hastily from a few facts and wins assent with phantom evidence alone. The antidote he prescribed was the table of absence and exclusion, whose whole point is the negative instance. That is precisely the operation of empirically measuring and lowering the denominator $\PrNC$. Bacon himself had no vocabulary of probability, and the \textit{three tables} and exclusion constituted a procedure of qualitative elimination rather than quantitative estimation. His \textit{table of exclusion} was left unfinished because it demanded a complete inventory of nature, yet the stance of seeking evidence on the side of absence returns as today's demand that the denominator be measured.

Generative AI has inflated this anticipation to an industrial scale. Of the four Idols Bacon listed, two are innate to the intellect (the Tribe and the Cave) and two are instilled from outside it: the Idols of the Marketplace enter through language and human intercourse, and those of the Theatre through received doctrines. What is new is neither the externality of the source nor its social character, but its industrialization: a device now produces, at near-zero marginal cost and on demand, the plausible-seeming artifacts that the Marketplace and the Theatre once supplied slowly and by accident. Let us call this extension the Idols of the Machine (\textit{Idola Machinae}): not so much a fifth kind as the Idols of the Marketplace and the Theatre grown to industrial scale. But naming is not curing. Just as Bacon's enumeration of the idols called for the table of absence, this name calls for the same procedure: measure the denominator. What Bacon sought to exclude is what statistics today treats as the denominator of the likelihood ratio, and the overestimate that arises when it is mismeasured is what we have named phantom evidence.

Phantom evidence is therefore not a disease peculiar to generative AI. It is the old habit of letting persuasiveness stand in for evidence, laid bare by the new condition that persuasiveness has become cheap. The prescription, likewise, is not peculiar to generative AI. The question to ask is not what can be produced but whether what has been produced can be shown not to have arisen by chance. Bacon's demand that evidence be sought on the side of absence was the first formulation of that question. Four centuries on, the question still has the same shape.

\clearpage
\appendix
\section*{Supplementary Material}
\noindent The main claim, that evidence lives in the denominator, stands complete without the appendices. Four supplements follow: the extreme-value origin of the selection term $\log m_{\mathrm{eff}}$ (Appendix~\ref{sec:extreme}), the subadditivity of positive cases (Appendix~\ref{sec:subadditivity}), a continuous version (Appendix~\ref{sec:continuous}) and the time axis (Appendix~\ref{sec:lifetime}).

\section{The Extreme-Value Law of Selection: Why ``Seek and Ye Shall Find''}\label{sec:extreme}
The selection term $\log m_{\mathrm{eff}}$ of the main text (Section~\ref{sec:mechanisms}) follows directly from extreme-value statistics. Under the null (with the target absent), let the statistics obtained from effectively $m$ independent analysis paths (random seeds, hyperparameters, preprocessing, metrics) be $Z_1,\dots,Z_m\overset{\text{i.i.d.}}{\sim}\mathcal N(0,1)$, and suppose only the best one, $M_m=\max_i Z_i$, is reported. Since the best being at most $t$ is the same as all $m$ being at most $t$,
\begin{equation}
\Pr(M_m\le t)=\Phi(t)^m
\end{equation}
where $\Phi$ is the standard normal cumulative distribution function. This is nothing other than $1-(1-\gamma)^{m}$ of Section~\ref{sec:mechanisms}, with the bare denominator read as the upper-tail probability $\gamma=1-\Phi(t)$. The typical size of the best can be measured by the upper $1/m$ point $t^\ast$, defined by $1-\Phi(t^\ast)=1/m$, and the Gaussian tail asymptotics give
\begin{equation}
t^\ast=\sqrt{2\ln m}-\frac{\ln\ln m+\ln 4\pi}{2\sqrt{2\ln m}}+o(1).
\end{equation}
Thus $\sqrt{2\ln m}$ is only the leading term and overshoots at practical $m$. The exact values $\Phi^{-1}(1-1/m)$ for $m=10^2,10^3,10^4,10^6$ are about $2.3,\,3.1,\,3.7,\,4.8\,\sigma$ (the leading term alone would give $3.0,\,3.7,\,4.3,\,5.3$, too large by $0.5$ to $0.7\,\sigma$), and a hit at the ``$0.1\%$ by chance'' level appears without difficulty at around $m=10^3$. That the growth is as slow as $\sqrt{\ln m}$ is no consolation, because what generative AI makes cheap is precisely the order of magnitude of $m$. To evaluate fairly one must raise the threshold to $t^\ast$, that is, demand $\log m_{\mathrm{eff}}$ more evidence, a logarithmic version of multiple-comparison correction; and the amount by which the threshold is not raised is exactly the selection term of phantom evidence.

\section{Subadditivity of Positive Cases: Correlated Hits}\label{sec:subadditivity}
Section~\ref{sec:lr} stated that for independent observations the weights of evidence add, so that $n$ hits give $W=\sum_i\log\LR_i=n\,w$. As a minimal model, suppose the statistic of each trial has equicorrelation $\varrho$ about a shared signal $\mu$, with $\varrho\in(-1/(n-1),1)$ and, in the cases of interest, $\varrho\ge0$. (The $\varrho$ of this appendix is a correlation coefficient and is distinct from the prior odds $\rho$ of Section~\ref{sec:ppv}.) The covariance is then $\Sigma=\sigma^2\{(1-\varrho)I+\varrho\mathbf{1}\mathbf{1}^{\!\top}\}$. Its eigenvalues are $\sigma^2\{1+(n-1)\varrho\}$ and $\sigma^2(1-\varrho)$, so positive definiteness is guaranteed for $-1/(n-1)<\varrho<1$. Counting the Fisher information about $\mu$,
\begin{equation}
I(\mu)=\mathbf{1}^{\!\top}\Sigma^{-1}\mathbf{1}
=\frac{n}{\sigma^{2}\{1+(n-1)\varrho\}}
=\frac{n_{\mathrm{eff}}}{\sigma^{2}},
\qquad
n_{\mathrm{eff}}=\frac{n}{1+(n-1)\varrho}
\end{equation}
so the information is proportional not to the count $n$ but to the effective count $n_{\mathrm{eff}}$. At $\varrho=0$ (independence) we recover additivity with $n_{\mathrm{eff}}=n$ (for $\varrho<0$ one would obtain $n_{\mathrm{eff}}>n$, superadditivity, which is outside our concern), and in the limit $\varrho\to1$ (complete sharing) $n_{\mathrm{eff}}\to1$: however many are lined up, they are worth one. In general $n_{\mathrm{eff}}\to1/\varrho$ saturates as $n\to\infty$, so for instance $\varrho=0.1$ caps it at about ten, and whether one runs $100$ or $1000$ trials the effective number is about ten. For the test of a mean shift ($H_1:\mu=\delta$ against $H_0:\mu=0$ with $\Sigma$ known), the expected weight of evidence is $\mathbb E[\log\LR]=D_{\mathrm{KL}}=(\delta^2/2)\,\mathbf{1}^{\!\top}\Sigma^{-1}\mathbf{1}=(\delta^2/2\sigma^2)\,n_{\mathrm{eff}}$, so in this setting the subadditivity of Fisher information carries over directly into a subadditivity of evidence.\footnote{In general the two are different quantities. A $\log\LR$ computed from the correct joint likelihood takes a single value even under correlation; what breaks is the naive sum of the individual ratios, $\sum_i\log\LR_i$. Here $n_{\mathrm{eff}}$ is an approximation that indicates the scale of that overestimate, and the correspondence fails once one departs from normality, known $\Sigma$, a local mean shift, or measurement in expectation.} This is another expression of the same story that $k_{\mathrm{eff}}$ tells: a gallery of positive cases from the same generator and the same analysis carries, however impressive, only $n_{\mathrm{eff}}$ cases' worth of evidence. So long as the reporting format takes the number of positive cases as the primary metric (Section~\ref{sec:prescription}(4)), this saturation remains invisible to the reader.

\section{Continuous Formulation: From Discrete Candidates to Density}\label{sec:continuous}
The discrete quantities of the main text, nominal $N$ and effective $k_{\mathrm{eff}}$, are the shadow of a discretization by resolution $\tau$. Let the output space be a region of volume $V$ and dimension $D$, let $u$ be the uniform density on it, and let $\tau$ be the smallest scale the measurement can resolve. In the continuous version, phantom evidence appears more naturally as the expectation of a log density ratio. Replacing counts with densities, $N=V/\tau^{D}$ (output-space volume $V$, dimension $D$), $k_{\mathrm{eff}}(\tau)$ is the $\tau$-covering number of $p_g$, and phantom evidence in the $\neg T$ world (averaged over $p_g$) is
\begin{equation}
\Delta_g=D_{\mathrm{KL}}(p_g\|u)
\end{equation}
where the differential entropy alone is coordinate-dependent but the KL is invariant. The keystone is output dimension collapse. When $p_g$ concentrates on an effective dimension $d_{\mathrm{eff}}\ll D$, the $\tau$-regularized differential entropy $h_D(\tau)=h_{d_{\mathrm{eff}}}+(D-d_{\mathrm{eff}})\log\tau$ gives
\begin{equation}
\Delta_g=(D-d_{\mathrm{eff}})\log(1/\tau)+\text{const},
\end{equation}
that is, phantom evidence reduces to the number of collapsed transverse dimensions times the logarithm of the resolution. Here $\tau$ is not arbitrary: measurement noise imposes a lower bound, and marginal dimensions with SNR $<1$ do not enter $d_{\mathrm{eff}}$. This is the physical reason why high resolution does not add evidence. For an ideal observer who sees all outputs, the expected evidence is bounded by $D_{\mathrm{KL}}(p_T\|p_g)$, how far the true output distribution lies from the generator's prior, and coarse-graining to the judgment $C$ can only reduce it, by the data-processing inequality. Note that this bounds an expectation over $p_T$, not the weight of evidence carried by a single observation. After coarse-graining, the expected evidence in the judgment channel is $D\big(\mathrm{Bern}(\PrC)\,\|\,\mathrm{Bern}(\gamma)\big)$, which rises to the main text's bound $\log k_{\mathrm{eff}}$ as $\PrC\to1$ and falls strictly below it otherwise. Note that the shape of the quantity changes with the distribution over which one averages: on $p_g$ it is the KL ($\Delta_g$), but in the $T$ world (over $p_T$) it is a difference of KLs, $\Delta_T=\mathbb E_{p_T}[\log(p_g/u)]$, which is not necessarily non-negative.

\section{The Time Axis: Decay and Lifetime of Phantom Evidence}\label{sec:lifetime}
In the main text the selection multiplicity $m_{\mathrm{eff}}$ was treated statically. Making time $t$ an explicit variable reveals that phantom evidence has a lifetime.

Two clocks. Separate the researcher's clock (the rate $\nu_R$ at which trials, peeks and selections accumulate) from the community's clock (the rate $\nu_C$ at which independent verifications and negative controls accumulate). The researcher-side denominator, with the leakage-inclusive denominator rate $\gamma_\lambda$, is $p_R(t)\approx 1-(1-\gamma_\lambda)^{\nu_R t}$, with saturation time constant $\tau_R=1/(\nu_R\gamma_\lambda)$ and half-saturation time $t_{R,1/2}\approx 0.69/(\nu_R\gamma_\lambda)$, the time at which the denominator reaches $1/2$, which, in the absence of leakage ($\gamma_\lambda=\gamma=1/k_{\mathrm{eff}}$), reduces to $0.69\,k_{\mathrm{eff}}/\nu_R$. Repeatedly peeking at a fixed threshold inflates the family-wise false-positive rate, and because successive peeks are correlated the inflation is not the naive $1-(1-\alpha)^n$. Under the null, the upward excursion grows on the scale of $\sqrt{2\ln\ln t}$, by the law of the iterated logarithm. The correct brake is to design in advance a boundary that controls the overall error rate including the times at which one looks, as in group-sequential methods, $\alpha$-spending and confidence sequences~\autocite{pocock1977,obrien1979,landemets1983,johari2022}.

Erosion after publication. The phantom evidence at publication time $t_0$ is $\Delta_{\mathrm{total}}(t_0)=\log\!\big(N\,p_R(t_0)\big)$. Thereafter, as independent verifications and negative controls accumulate in the community, the illusion relaxes. We posit the simplest functional form, exponential relaxation, $\Delta_C(t)=\Delta_{\mathrm{total}}(t_0)\,e^{-(t-t_0)/\tau_C}$ with half-life $t_{C,1/2}=\tau_C\ln2$; it is an assumption, not a derived law. The decline effect, in which effect sizes shrink in replications~\autocite{schooler2011}, can be read as this decrease. The order of magnitude of $\tau_C$ is suggested by the ``half-life of truth''~\autocite{poynard2002} and the refutation of highly cited studies~\autocite{ioannidis2005jama}, on the scale of years to decades; the replication rates of large-scale replication studies~\autocite{osc2015,camerer2018} are not direct estimates of $\tau_C$ but auxiliary evidence that community correction is substantial. The disappearance of negative results~\autocite{fanelli2011} lowers $\nu_C$ and acts as a pressure that lengthens the lifetime.

\begin{figure}[t]
\centering
\input{figs/fig_time_en.tikz}
\caption{The lifetime of evidence: two clocks. During the research phase ($t\le t_0$), denominator inflation makes the effective evidence $\log_2\Lambda_{\mathrm{actual}}$ (blue) collapse quickly, but the observer's perception $\log_2\Lambda_{\mathrm{perceived}}$ (red) stays high. After publication, the illusion relaxes exponentially and slowly through independent verification and negative controls (time constant $\tau_C$). The shaded area is phantom evidence $\Delta_{\mathrm{total}}(t)$. The setting adds $\nu_R=10$/day and $\tau_C=10$ years to the running example of Section~\ref{sec:phantom}. The research phase and the post-publication phase differ by three orders of magnitude in time scale, so the axis is broken into two panels, each drawn in its own real units (days, years). The blue dashed line is the ceiling on evidence, $\log_2 k_{\mathrm{eff}}$, from which $\Lambda_{\mathrm{actual}}$ begins to collapse during the research phase.}
\label{fig:time}
\end{figure}
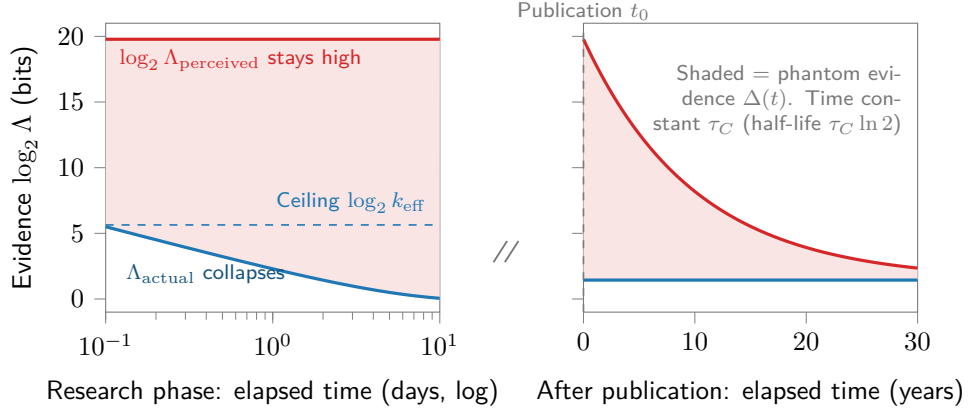

Lifetime and the mechanization of asymmetry. The lifetime is governed by the ratio of time constants $\eta=\tau_C/\tau_R$, with $\eta\gg1$ meaning that apparent discoveries can be made quickly while the verification that overturns them is slow. Generative AI raises $\nu_R$ and so lowers $\tau_R$, while $\nu_C$, which depends on independent data, independent teams and human scrutiny, does not accelerate at the same rate. Hence $\eta\gg1$: a device that speeds up the clock that makes phantoms and relatively slows the clock that breaks them. This is the main text's diagnosis, the industrialization of persuasiveness, extended to the time axis (Figure~\ref{fig:time}).

Distribution and inventory (Little's law). Verifiability is heterogeneous across findings, so the lifetime is a mixture of exponential distributions with a heavy tail, a residual fraction $p_\infty$ that is never verified, the standing stock of ``zombie findings.'' The inventory of live phantom evidence accumulating in a field is, by Little's law of queueing theory~\autocite{little1961},
\begin{equation}
\mathcal L=\nu_{\mathrm{pub}}\,\mathbb E[\Delta\,T_{\mathrm{life}}]\ \approx\ \nu_{\mathrm{pub}}\cdot\bar\Delta\cdot\bar\tau_C
\end{equation}
Here $\nu_{\mathrm{pub}}$ is the rate of publications containing phantoms, $\bar\Delta$ the average phantom per item and $\bar\tau_C$ the average lifetime. Factorizing into a product is an approximation that ignores the correlation between $\Delta$ and the lifetime $T_{\mathrm{life}}$. Generative AI moves all three factors in the bad direction. The publication rate $\nu_{\mathrm{pub}}$ rises, the phantom per item $\bar\Delta$ rises (cheap $m_{\mathrm{eff}}$ and a huge $N$), and verification is not accelerated so $\bar\tau_C$ is unchanged or increasing. The inventory therefore swells multiplicatively. The prescription (Section~\ref{sec:prescription}) must act on all three factors: the inflow $\nu_{\mathrm{pub}}$, the residence time $\bar\tau_C$, and the source $\bar\Delta$. Only the last two are addressed here; lowering the inflow of phantom-bearing publications is a matter of editorial policy rather than of measurement.

\printbibliography[title={References}]
\end{document}

%% file: figs/figstyle.tex
\newcommand{\figfont}{\sffamily}
\ifdefined\gtfamily\renewcommand{\figfont}{\sffamily\gtfamily}\fi
\tikzset{every picture/.append style={/utils/exec=\figfont}}

\definecolor{cGenuine}{RGB}{31,119,180}   
\definecolor{cDenom}{RGB}{214,39,40}      
\definecolor{cDenom2}{RGB}{255,140,0}     
\definecolor{cNeutral}{RGB}{120,120,120}  

\pgfplotsset{
  fable/.style={
    tick align=outside, tick pos=left,
    axis line style={cNeutral},
    every axis plot/.append style={thick},
    label style={font=\small},
    tick label style={font=\small},
    legend style={font=\small, draw=none, fill=none, legend cell align=left},
  }
}

\tikzset{
  fbox/.style={rounded corners=2pt, draw=cNeutral, line width=0.6pt, inner sep=5pt, align=center, font=\small},
  fgood/.style={fbox, draw=cGenuine, fill=cGenuine!8},
  fbad/.style={fbox, draw=cDenom, fill=cDenom!8},
  farrow/.style={-{Stealth[length=5pt]}, cNeutral, line width=0.8pt},
}

%% file: figs/fig_threecases_en.tikz
\resizebox{\linewidth}{!}{%
\begin{tikzpicture}[font=\footnotesize]
  \node[font=\scriptsize\itshape, text=cNeutral] at (1.9,3.5) {Nominal space $N$};
  \node[font=\scriptsize\itshape, text=cNeutral] at (7.5,3.5) {Effective options $k_{\mathrm{eff}}$};
  \foreach \i/\y/\name/\nom/\eff/\mech in {%
     A/2.6/{Clever Hans}/{All responses}/{The cue subspace}/{Unconscious\\cues},%
     B/1.3/{Overrated\\LLM ability}/{All token\\sequences}/{A narrow solution set}/{Memorization,\\contamination},%
     C/0/{Spurious\\reconstruction}/{All images}/{A few dozen modes}/{Generative prior\\{}+ classification}}{%
     \node[fbox, anchor=west, text width=30mm, font=\footnotesize, minimum height=9mm] (n\i) at (0.6,\y) {\nom};
     \node[fgood, anchor=west, text width=17mm, font=\scriptsize] (k\i) at (6.6,\y) {\eff};
     \draw[farrow] (n\i.east) -- (k\i.west);
     \node[above, font=\scriptsize, text=cNeutral, align=center, inner sep=1pt]
       at ($(n\i.east)!0.5!(k\i.west)$) {\mech};
     \node[anchor=east, font=\scriptsize, text=cNeutral, align=right, text width=27mm] at (0.4,\y) {\name};
  }
  \node[fbad, anchor=west, text width=24mm, font=\footnotesize] (d) at (9.9,1.3)
    {Phantom evidence\\[1pt]$\Delta=\log\dfrac{N}{k_{\mathrm{eff}}}$};
  \draw[farrow] (kA.east) -- (d.north west);
  \draw[farrow] (kB.east) -- (d.west);
  \draw[farrow] (kC.east) -- (d.south west);
\end{tikzpicture}%
}

%% file: figs/fig_boxes_en.tikz
\begin{tikzpicture}[font=\small]
  \node[anchor=south west, font=\footnotesize, cNeutral, align=left] at (0,3.20)
    {Observer's assumption:\\nominal $N=10^6$ (faint whole space)};
  \foreach \i in {0,...,11}{\foreach \j in {0,...,8}{
    \fill[cNeutral!12] (\i*0.4,\j*0.34) rectangle ++(0.32,0.26);}}
  \draw[cGenuine, line width=0.8pt] (8*0.4-0.05,4*0.34-0.05) rectangle ++(0.42,0.36);
  \fill[cDenom] (8*0.4+0.09,4*0.34+0.07) rectangle ++(0.14,0.12);
  \node[anchor=north, font=\footnotesize, cDenom] at (2.3,-0.15)
    {Felt as ``$1/N$ by chance''};
  \begin{scope}[shift={(7.2,0.30)}]
    \node[anchor=south west, font=\footnotesize, cNeutral, align=left] at (-0.12,2.90)
      {What the system actually reaches:\\effective $k_{\mathrm{eff}}=50$};
    \foreach \i in {0,...,4}{\foreach \j in {0,...,3}{
      \fill[cGenuine!20, draw=cGenuine!55] (\i*0.8,\j*0.66) rectangle ++(0.66,0.54);}}
    \fill[cDenom] (2*0.8,2*0.66) rectangle ++(0.66,0.54);
    \draw[cGenuine, line width=0.8pt] (-0.12,-0.12) rectangle (4.10,2.78);
    \node[anchor=north, font=\footnotesize, cDenom] at (1.9,-0.5)
      {Should be counted as $1/k_{\mathrm{eff}}$};
  \end{scope}
  \draw[cGenuine!70, line width=0.5pt, dashed] (8*0.4+0.37,4*0.34+0.31) -- (7.08,3.08);
  \draw[cGenuine!70, line width=0.5pt, dashed] (8*0.4+0.37,4*0.34-0.05) -- (7.08,0.18);
  \node[font=\scriptsize, cGenuine!70!black, align=center] at (5.75,1.66)
    {One cell of the same space,\\magnified $\times\,2\times10^{4}$};
\end{tikzpicture}

%% file: figs/fig_delta_en.tikz
\begin{tikzpicture}[font=\small,
   lab/.style={anchor=north, font=\scriptsize, text=cNeutral, align=center, text width=2.25cm}]
  \def\s{0.30}
  \def\w{1.30}
  \draw[cNeutral] (0,0) -- (0,20.4*\s);
  \foreach \b in {0,5,10,15,20}{\draw[cNeutral] (-0.09,\b*\s) -- (0,\b*\s)
      node[left, font=\scriptsize, cNeutral] {\b};}
  \node[rotate=90, anchor=south, font=\scriptsize, cNeutral] at (-0.78,10*\s) {$\log_2\Lambda$ (bits)};
  \draw[cNeutral!60, dashed] (0,0) -- (9.4,0);
  \fill[cDenom!25, draw=cDenom, line width=0.8pt] (0.6,0) rectangle ++(\w,19.78*\s);
  \node[anchor=south, font=\scriptsize, cDenom] at (0.6+\w/2,19.78*\s+0.06) {$19.8$};
  \node[lab] at (0.6+\w/2,-0.14) {Perceived evidence\\$\log_2\Lambda_{\mathrm{perceived}}$};
  \fill[cDenom!45, draw=cDenom, line width=0.8pt] (3.0,5.49*\s) rectangle ++(\w,14.29*\s);
  \draw[cNeutral!60, dashed, line width=0.5pt] (1.90,19.78*\s) -- (3.0,19.78*\s);
  \node[font=\scriptsize, cDenom!70!black] at (3.0+\w/2,12.6*\s) {$-14.3$};
  \node[lab] at (3.0+\w/2,-0.14) {Nominal vs.\ effective\\$\log(N/k_{\mathrm{eff}})$};
  \fill[cDenom!45, draw=cDenom, line width=0.8pt] (5.4,1.44*\s) rectangle ++(\w,4.05*\s);
  \draw[cDenom!70, dashed, line width=0.6pt] (5.4,1.17*\s) rectangle ++(\w,4.32*\s);
  \draw[cNeutral!60, dashed, line width=0.5pt] (4.30,5.49*\s) -- (5.4,5.49*\s);
  \node[anchor=west, font=\scriptsize, cDenom!70!black] at (5.4+\w+0.06,3.4*\s) {$-4.1$};
  \node[lab] at (5.4+\w/2,-0.14) {Selection $\log m_{\mathrm{eff}}$\\(dashed $=$ additive approx.\ $-4.3$)};
  \fill[cGenuine!30, draw=cGenuine, line width=0.8pt] (7.8,0) rectangle ++(\w,1.44*\s);
  \draw[cNeutral!60, dashed, line width=0.5pt] (6.70,1.44*\s) -- (7.8,1.44*\s);
  \node[anchor=south, font=\scriptsize, cGenuine!70!black] at (7.8+\w/2,1.44*\s+0.06) {$1.4$};
  \node[lab] at (7.8+\w/2,-0.14) {Actual evidence\\$\log_2\Lambda_{\mathrm{actual}}$};
  \draw[{Stealth[length=4pt]}-{Stealth[length=4pt]}, cDenom!70!black, line width=0.9pt]
    (9.75,1.44*\s) -- (9.75,19.78*\s);
  \node[rotate=90, anchor=north, font=\scriptsize, cDenom!70!black] at (9.68,10.6*\s)
    {Phantom evidence $\Delta_{\mathrm{total}}=18.3$ bits};
  \node[anchor=north west, font=\scriptsize, text=cNeutral, align=left] at (0.1,-1.95)
    {The third term, leakage $\log(\gamma_\lambda/\gamma)$, is $0$ in the running example ($\lambda=1$); it can be large in real data.};
\end{tikzpicture}

%% file: figs/fig_ppvladder_en.tikz
%
\resizebox{\linewidth}{!}{%
\begin{tikzpicture}[font=\small]
  \def\s{0.28}
  \draw[cNeutral] (0,0) -- (0,20.8*\s);
  \foreach \b in {0,5,10,15,20}{\draw[cNeutral] (-0.09,\b*\s) -- (0,\b*\s)
      node[left, font=\scriptsize, cNeutral] {\b};}
  \node[rotate=90, anchor=south, font=\scriptsize, cNeutral] at (-0.80,10*\s) {$\log_2\Lambda$ (bits)};
  \draw[cNeutral, line width=0.8pt] (0,0) -- (8.9,0);
  \node[anchor=north west, font=\scriptsize, cNeutral] at (0,-0.10)
    {Common footing: prior $\pi=\Pr(T)$. $\log_2\Lambda=0$ is the death of evidence};
  \draw[cDenom, dashed, line width=0.7pt] (0.45,19.93*\s) -- (8.6,19.93*\s);
  \node[anchor=south east, font=\scriptsize, cDenom] at (8.6,19.93*\s+0.05)
    {Observer's ceiling $=19.9$};
  \draw[cGenuine, dashed, line width=0.7pt] (0.45,5.64*\s) -- (8.6,5.64*\s);
  \node[anchor=south east, font=\scriptsize, cGenuine] at (8.6,5.64*\s+0.05)
    {Ceiling $\log_2 k_{\mathrm{eff}}=5.6$};
  \draw[cDenom, line width=1.1pt] (1.65,0) -- (1.65,19.78*\s);
  \draw[cDenom, line width=1.1pt] (2.35,0) -- (2.35,19.78*\s);
  \foreach \b in {1,...,19}{\draw[cDenom, line width=0.5pt] (1.65,\b*\s) -- (2.35,\b*\s);}
  \node[anchor=south, cDenom, font=\scriptsize, align=center] at (2,19.78*\s+0.55)
    {perceived $\log_2\Lambda=19.8$};
  \draw[cGenuine, line width=1.1pt] (4.15,0) -- (4.15,5.49*\s);
  \draw[cGenuine, line width=1.1pt] (4.85,0) -- (4.85,5.49*\s);
  \foreach \b in {1,...,5}{\draw[cGenuine, line width=0.5pt] (4.15,\b*\s) -- (4.85,\b*\s);}
  \node[anchor=south, cGenuine!70!black, font=\scriptsize, align=center] at (4.5,5.49*\s+0.42)
    {actual (concentration only) $=5.5$};
  \draw[cGenuine!55, line width=1.1pt] (6.65,0) -- (6.65,1.44*\s);
  \draw[cGenuine!55, line width=1.1pt] (7.35,0) -- (7.35,1.44*\s);
  \draw[cGenuine!55, line width=0.5pt] (6.65,1*\s) -- (7.35,1*\s);
  \node[anchor=west, cGenuine!60!black, font=\scriptsize, align=left] at (7.45,1.44*\s*0.6)
    {Actual (with selection) $=1.4$};
  \draw[{Stealth[length=4pt]}-{Stealth[length=4pt]}, cNeutral, line width=0.7pt]
    (3.25,5.49*\s) -- (3.25,19.78*\s);
  \node[anchor=west, font=\scriptsize, cNeutral] at (3.38,17.3*\s)
    {Nominal vs.\ effective confusion $14.3$};
  \draw[{Stealth[length=4pt]}-{Stealth[length=4pt]}, cNeutral, line width=0.7pt]
    (5.75,1.44*\s) -- (5.75,5.49*\s);
  \node[anchor=west, font=\scriptsize, cNeutral] at (5.88,3.6*\s)
    {Selection $4.1$ (additive approx.\ $4.3$)};
  \draw[{Stealth[length=4pt]}-{Stealth[length=4pt]}, cDenom!70!black, line width=0.9pt]
    (0.85,1.44*\s) -- (0.85,19.78*\s);
  \node[rotate=90, anchor=south, font=\scriptsize, cDenom!70!black] at (-1.30,10.6*\s)
    {Phantom evidence $\Delta_{\mathrm{total}}=18.3$};
  \draw[-{Stealth[length=5pt]}, cGenuine, line width=1.1pt] (9.35,5.64*\s) -- (9.35,19.6*\s);
  \node[rotate=90, anchor=south, font=\scriptsize, cGenuine, align=center] at (9.50,12.6*\s)
    {Direction 1: raise the ceiling\\($k_{\mathrm{eff}}\to N$)};
  \draw[-{Stealth[length=5pt]}, cDenom, line width=1.1pt] (10.9,19.6*\s) -- (10.9,1.8*\s);
  \node[rotate=90, anchor=north, font=\scriptsize, cDenom, align=center] at (10.75,10.6*\s)
    {Direction 2: pull $\log N$ down\\to the measured denominator};
\end{tikzpicture}%
}

%% file: figs/fig_negcontrol_en.tikz
\resizebox{\linewidth}{!}{%
\begin{tikzpicture}[font=\small,
    nb/.style={fbox, text width=4.0cm, minimum height=1.45cm, align=center},
    rl/.style={font=\footnotesize\bfseries, align=right, anchor=east},
  ]
  \node[nb, fbad] (A1) at (0,0)   {Stop only the generator};
  \node[nb, fbad] (A2) at (4.9,0) {Measure only bare $\gamma\approx1/k_{\mathrm{eff}}$\\[1pt]
    {\scriptsize (running ex.: $k_{\mathrm{eff}}^{\mathrm{obs}}=50$)}};
  \node[nb, fbad] (A3) at (9.8,0) {Denominator underestimated\\[1pt]
    {\scriptsize Reported as $\log_2\Lambda=5.5$ bits}};
  \draw[farrow, cDenom!70] (A1) -- (A2);
  \draw[farrow, cDenom!70] (A2) -- (A3);
  \node[rl, cDenom] at (-2.30,0) {Shallow\\control};
  \node[anchor=north, font=\scriptsize, cDenom] at (4.9,-1.15)
    {Misses selection $m_{\mathrm{eff}}$ and leakage $\lambda$};
  \node[nb, fgood] (B1) at (0,-3.0)   {Same pipeline and selection rule on independent data};
  \node[nb, fgood] (B2) at (4.9,-3.0) {Measure effective $\gamma_\lambda$ (incl.\ $m_{\mathrm{eff}},\lambda$)\\[1pt]
    {\scriptsize (running ex.: $k_{\mathrm{eff}}^{\mathrm{obs}}=3.0$)}};
  \node[nb, fgood] (B3) at (9.8,-3.0) {Denominator measured correctly\\[1pt]
    {\scriptsize $\log_2\Lambda=1.4$ bits}};
  \draw[farrow, cGenuine!70] (B1) -- (B2);
  \draw[farrow, cGenuine!70] (B2) -- (B3);
  \node[rl, cGenuine] at (-2.30,-3.0) {Pipeline-\\complete};
  \node[anchor=north, font=\scriptsize, cGenuine!70!black] at (4.9,-4.15)
    {Gives $\lambda\to0$ and measures $m_{\mathrm{eff}}$};
\end{tikzpicture}%
}

%% file: figs/fig_time_en.tikz
\begin{tikzpicture}
\begin{axis}[fable, name=L, width=6.0cm, height=5.4cm,
  xmode=log, xmin=0.1, xmax=10, ymin=-1, ymax=21,
  xlabel={Research phase: elapsed time (days, log)},
  xlabel style={at={(axis description cs:0.5,0)}, anchor=north, yshift=-7.5mm},
  ylabel={Evidence $\log_2\Lambda$ (bits)},
  ytick={0,5,10,15,20}, clip=false]
\addplot[name path=pL, cDenom, line width=1.2pt, domain=0.1:10, samples=2] {19.78};
\addplot[name path=aL, cGenuine, line width=1.2pt, domain=0.1:10, samples=200]
  {max(-0.5, log2(0.9) - log2(1-pow(0.98,10*x)))};
\addplot[cDenom!12] fill between[of=pL and aL];
\draw[cGenuine, dashed, line width=0.6pt] (axis cs:0.1,5.64) -- (axis cs:10,5.64);
\node[anchor=south east, font=\scriptsize, cGenuine] at (axis cs:9.5,5.80) {Ceiling $\log_2 k_{\mathrm{eff}}$};
\node[anchor=west, font=\scriptsize, cDenom] at (axis cs:0.105,18.2) {$\log_2\Lambda_{\mathrm{perceived}}$ stays high};
\node[anchor=south west, font=\scriptsize, cGenuine!70!black] at (axis cs:0.115,0.4) {$\Lambda_{\mathrm{actual}}$ collapses};
\end{axis}
\draw[cNeutral, line width=0.7pt] ([xshift=8mm,yshift=8mm]L.south east) ++(-0.09,-0.14) -- ++(0.18,0.28);
\draw[cNeutral, line width=0.7pt] ([xshift=9.5mm,yshift=8mm]L.south east) ++(-0.09,-0.14) -- ++(0.18,0.28);
\begin{axis}[fable, name=R, at={(L.east)}, xshift=1.9cm, anchor=west,
  width=6.0cm, height=5.4cm,
  xmin=0, xmax=30, ymin=-1, ymax=21,
  xlabel={After publication: elapsed time (years)},
  xlabel style={at={(axis description cs:0.5,0)}, anchor=north, yshift=-7.5mm},
  ytick={0,5,10,15,20}, yticklabels={,,,,},
  clip=false]
\addplot[name path=pR, cDenom, line width=1.2pt, domain=0:30, samples=200] {1.44+18.34*exp(-x/10)};
\addplot[name path=aR, cGenuine, line width=1.2pt, domain=0:30, samples=2] {1.44};
\addplot[cDenom!12] fill between[of=pR and aR];
\node[anchor=north east, font=\scriptsize, cNeutral, align=right, text width=3.9cm] at (axis cs:29.5,18.6)
  {Shaded $=$ phantom evidence $\Delta(t)$. Time constant $\tau_C$ (half-life $\tau_C\ln 2$)};
\draw[cNeutral, dashed, line width=0.6pt] (axis cs:0,-1) -- (axis cs:0,20.4);
\node[anchor=south, font=\scriptsize, cNeutral] at (axis cs:0,20.5) {Publication $t_0$};
\end{axis}
\end{tikzpicture}